\documentclass{article}
\usepackage{graphicx} 
\usepackage{xcolor}
\usepackage{hyperref}
\usepackage[labelfont=bf]{caption}
\usepackage{booktabs}
\usepackage{amsmath, amsfonts, amssymb, textcomp}
\usepackage{float}
\usepackage{authblk}
\usepackage[utf8]{inputenc}
\usepackage[T1]{fontenc}

\newfloat{suppfigure}{tbp}{losf}
\floatname{suppfigure}{Supplementary Figure}

\title{An imageless magnetic resonance framework for fast and cost-effective decision-making}
\author[1]{Alba González-Cebrián\thanks{These authors contributed equally}}
\author[1]{Pablo García-Cristóbal\textsuperscript{*}}
\author[1]{Fernando Galve}
\author[2]{Efe Ilıcak}
\author[2]{Viktor Van Der Valk}
\author[2]{Marius Staring}
\author[2]{Andrew Webb}
\author[1]{Joseba Alonso}

\affil[1]{Instituto de Instrumentación para Imagen Molecular, CSIC, 
          Universitat Politècnica de València}
\affil[2]{Leiden University Medical Center, Department of Radiology, 
          Leiden, The Netherlands}

\date{\today}

\begin{document}
	
	\maketitle
    \begin{abstract}
        Magnetic Resonance Imaging (MRI) is the gold standard in countless diagnostic procedures, yet hardware complexity, long scans, and cost preclude rapid screening and point-of-care use. We introduce \emph{Imageless Magnetic Resonance Diagnosis} (IMRD), a framework that bypasses $k$-space sampling and image reconstruction by analyzing raw one-dimensional MR signals. We identify potentially impactful embodiments where IMRD requires only optimized pulse sequences for time-domain contrast, minimal low-field hardware, and pattern recognition algorithms to answer clinical closed queries and quantify lesion burden. As a proof of concept, we simulate multiple sclerosis lesions in silico within brain phantoms and deploy two extremely fast protocols (approximately 3 s), with and without spatial information. A 1D convolutional neural network achieves AUC close to 0.95 for lesion detection and $R^2$ close to 0.99 for volume estimation. We also perform robustness tests under reduced signal-to-noise ratio, partial signal omission, and relaxation-time variability. By reframing MR signals as direct diagnostic metrics, IMRD paves the way for fast, low-cost MR screening and monitoring in resource-limited environments.
    \end{abstract}

	\section{Introduction}
	\label{sec:intro}
    Clinical diagnosis is a decision-making process combining various data sources to assign each patient to a diagnostic category predefined by the medical profession \cite{ball2015improving,jutel2009sociology}. Along this process, diagnostic tests pose closed questions to confirm or rule out differential diagnoses and determine the process sensitivity and precision. The accuracy and cost-effectiveness reached by some of these tests underpin screening protocols \cite{maxim2014screening}--conducted even in asymptomatic individuals--as well as patient monitoring during and after treatment \cite{european2015medical}. 
    
    In this context, Magnetic Resonance Imaging (MRI) is an attractive asset because it is both non-invasive and non-ionising \cite{brown2014magnetic}, offering high-contrast, high-resolution anatomical images as compelling evidence to detect abnormal tissue patterns and diagnose underlying pathologies \cite{doi2006diagnostic,hussain2022modern}. However, these high-fidelity images impose hardware and schedule requirements that severely hinder the utility of routine MRI in screening and follow-up, contributing to its high cost and limited accessibility \cite{ladd2009whole,schmidt2010uses,iglehart2006new,garrahy2024towards}. This scarce access to MR systems and bottlenecks associated with MR usage have motivated research on viable low-field (LF) scanners \cite{Cooley2020,OReilly2020,Guallart-Naval2022,Obungoloch2023,Webb2023}. These LF MRI scanners focus on implementing low-cost systems at the expense of image resolution, hindering human-sight recognition on the resulting images.
    
    Nonetheless, image-centred analysis perpetuates multiple constraints, inflating MR costs. Acquisition protocols devote most of their time to encoding spatial information for a later visual reconstruction, and these long scanning times are coupled with extremely sophisticated hardware that includes magnets for highly homogeneous EM fields, linear gradients and shieldings. All these elements are tied to the subsequent reconstruction of a faithful image from the acquired MR signals.
  
    This has motivated \textit{imageless} approaches, which decouple MR signals from image reconstruction \cite{du2024mri, patent2019,singhal2023feasibility}. Because Fourier transformations between frequency (i.e. $k$-space) and image domains are information-lossless, working on the $k$-space maintains signal fidelity, meaning that $k$-space data holds the same diagnostic value as reconstructed images. Thus, the same information is numerically there but differently represented, which can be exploited by non-human data-processing methods, namely AI models, applied to analyse such $k$-space data.  
    Some examples are systems proposed in \cite{patent2019} and \cite{singhal2023feasibility}, where an NMR scanner is coupled with AI to identify $k$-space subsets with the most discriminant power to calculate a risk score for prostate cancer. Still, these methods require relatively comprehensive $k$-space sampling and thus long scan times. By questioning the necessity of integrating abundant spatial information, more radical imageless solutions could reduce MR data requirements and expand MR viability in screening or diagnosis tests that do not rely on traditional images.

    In this work, we propose a fundamental shift in MR data acquisition and analysis, which we call Imageless Magnetic Resonance Diagnosis (IMRD), to allow for better MR integration as a tool for fast and cost-effective clinical decision-making. 
    Rather than relying on spatial-frequency data in two or more dimensions \cite{du2024mri,singhal2023feasibility}, IMRD operates directly on raw MR data (i.e. 1D signals in the time domain), bypassing $k$-space itself. The proposed imageless protocol involves (i) optimising an MR sequence that triggers a change in the MR signal if an event of interest is present in a sample, (ii) scanning compatible with minimal hardware, and (iii) a pattern recognition model to detect the presence of an event of interest. In the following, we first outline the IMRD framework, covering potential applications, its impact on lightening hardware requirements and its reliance on optimal information extraction and processing. Then, we illustrate an imageless use case with simulated MR signals from in silico white matter lesions \cite{da2019multiple}, serving as a toy model for a potential application of an imageless framework to answer a clinical question of interest (i.e., to detect and/or size lesions in this particular example). Next, we investigate how more realistic and less stable conditions could affect model performance in an imageless setting, and finally, we highlight the main conclusions and IMRD's potential limitations. 

    \section{IMRD framework}
    \label{sec:imrd_framework}
    The central hypothesis of the IMRD framework is that closed questions, common in screening and monitoring tasks, can be answered quickly and reliably with simplified MR hardware, optimised pulse sequences, and advanced data processing. In this section, we first discuss the types of relevant tests susceptible to IMRD methods and then move on to potential implications on scanner hardware, information encoding, and conclusion extraction. The overarching question guiding these discussions is: \emph{what is the absolute minimal structure required to address a relevant clinical inquiry?}

    \subsection{Potential applications}
    \label{sec:imrd_applications}    
    Classification and regression are well-suited tasks for automated decision-making based on AI, leveraging Deep Learning and advanced data processing. Potentially interesting tasks for IMRD include those that seek:
    \begin{itemize}
        \item the \emph{presence or absence} of a substance or tissue, as e.g. in white matter lesions, where tissue relaxation measurements can be used to diagnose a disease such as Multiple Sclerosis (MS) \cite{bakshi2008mri,arnold2022sensitivity};
        \item the \emph{abundance} of a substance or tissue, as e.g. in hydrocephalus, characterised by an abnormal build-up of cerebrospinal fluid (CSF) \cite{harper2021assessing}; 
        \item anomalies in the \emph{structural disposition or spatial distribution} of a tissue or an organ, as e.g. in lissencephaly, whereby parts of the brain surface appear smooth \cite{krawinkel1987magnetic};
        \item anomalies in the \emph{texture or stiffness} of a tissue or organ, as e.g. in liver fibrosis, where MR elastographic techniques are useful to gauge the severity of the disease \cite{venkatesh2013magnetic};
        \item an abnormal \emph{temperature} of a tissue or organ, as e.g. in critically ill neurologic patients, where MR thermometry can be employed to diagnose cerebrovascular diseases \cite{dehkharghani2020mr};
        \item specific responses to \emph{pulse sequences}, as e.g. with magnetization preparation modules \cite{Storey2015} or functional MR \cite{glover2011overview};
        \item specific responses to \emph{magnetic environments}, as e.g. in the presence of contrast agents such as those based on gadolinium \cite{caravan1999gadolinium}; or
        \item \emph{temporal changes} of a lesion or organ, as in patient monitoring or longitudinal studies \cite{ross2011review}. 
    \end{itemize}

    MR techniques are currently employed in all these tasks, albeit with arguably limited global impact due to the generally scarce access to scanners. In line with the IMRD paradigm of decision-making with minimal resources, we discuss how to tackle some of them by an imageless framework in the following subsections.

    \subsection{Impact of IMRD on hardware requirements}
    Clinical MR scanners impose formidable engineering constraints to produce human-readable images: the remarkable signal-to-noise ratio and spatial resolution demand superconducting magnets to produce extreme magnetic fields which are exceptionally uniform, avoiding confounding artifacts and spatial distortions, and the gradient coils must generate highly linear fields to arrange the received signals into a $k$-space that corresponds to the spatial frequency representation of the scanned subject. This far-from-exhaustive list already indicates that the limited access to MRI worldwide stems from tailoring MRI scanners to human perception, which is overwhelmingly visual.

    Previous imageless approaches \cite{du2024mri, patent2019,singhal2023feasibility} relied on scanners designed to meet the above specifications, aiming to balance $k$-space undersampling with properly answering a clinically relevant question, even if no meaningful image was reconstructed. IMRD dives deeper to acquire and analyse the electrical time-dependent signals fed into the scanner console, thereby operating in signal space directly and relaxing engineering constraints.

    One example application where IMRD can have a strong impact on hardware requirements could be hydrocephalus screening and/or monitoring. The excess in cerebrospinal fluid (CSF) imprints MR signals acquired during a pulse sequence, so even handheld devices (similar to those in \cite{eidmann1996nmr} or \cite{sherman2024single}) could be employed to estimate its abundance. Such a scanner could consist of a single-sided magnet, built-in inhomogeneities (i.e. no gradients), and a single radio-frequency (RF) coil tuned to be sensitive to different depths, all of which could be assembled for less than 3\,k€.

    Another example would be detecting white matter lesions, as in MS, for which $T_1$ and $T_2$ relaxation times differ from healthy ones. A simplified version of a head-only scanner \cite{sheth2021assessment,o2021vivo,galve2024elliptical} could gather relaxometry data to indicate and quantify the extent of damaged tissues. This is possible with a single gradient or even without gradients, provided the main magnetic field is homogeneous enough that $T_2^*$ and $T_2$ times do not differ greatly. Sections \ref{sec:method} and \ref{sec:results} are a deep dive into precisely this example and aim to validate the IMRD concept in silico.
        
    \subsection{Optimal information encoding and data processing}
    The optimal design of highly efficient electromagnetic (EM) pulse sequences to generate MR signals that maximise tissue discrimination for classification or regression tasks is a field on its own \cite{zhao2018optimal} and is likely to be dependent on the addressed clinical question, the MR hardware and operation, and the data processing pipeline. One approach in the IMRD context is to use physics-informed analytical models of MR signal response and seek for strictly optimal EM sequences that maximise a given contrast \cite{zhao2018optimal}. Alternatively, a data-driven process would run various pulse sequences on a controlled number of subjects, and let an AI find key discriminant patterns.
    
    In either case, optimal EM sequences may be more organic than standard concatenations of hard and trapezoidal pulses. This is unlikely to require steady-state magnetization, in stark contrast to conventional MRI pulse sequences, and could increase the useful duty cycle (i.e. the ratio between data acquisition and overall sequence durations). The EM pulse sequences employed in the remainder of the paper were inspired by Magnetic Resonance Fingerprinting, which also avoids steady-state magnetization \cite{ma2013magnetic}.

    Likewise, data processing strategies can range between simple \cite{sherman2024single} or complex \cite{sbrizzi2018fast} physics-informed analytic models, and purely data-driven approaches, where labelled or unlabelled data are fed to an AI to identify different conditions \cite{du2024mri}. In the following sections, we benchmark the performance of both approaches in the context of IMRD with simulated MS lesions.
        	
	\section{Methodology}
	\label{sec:method}
	The in-silico studies presented in this work comprise two primary elements: simulated datasets and models to estimate the volume and the presence of MS lesions. This section describes each component and the technical details of its implementation.
	\subsection*{Simulation of MR signals}
	\label{sec:method_sequence}
\begin{figure}
        \centering
        \includegraphics[width=\linewidth]{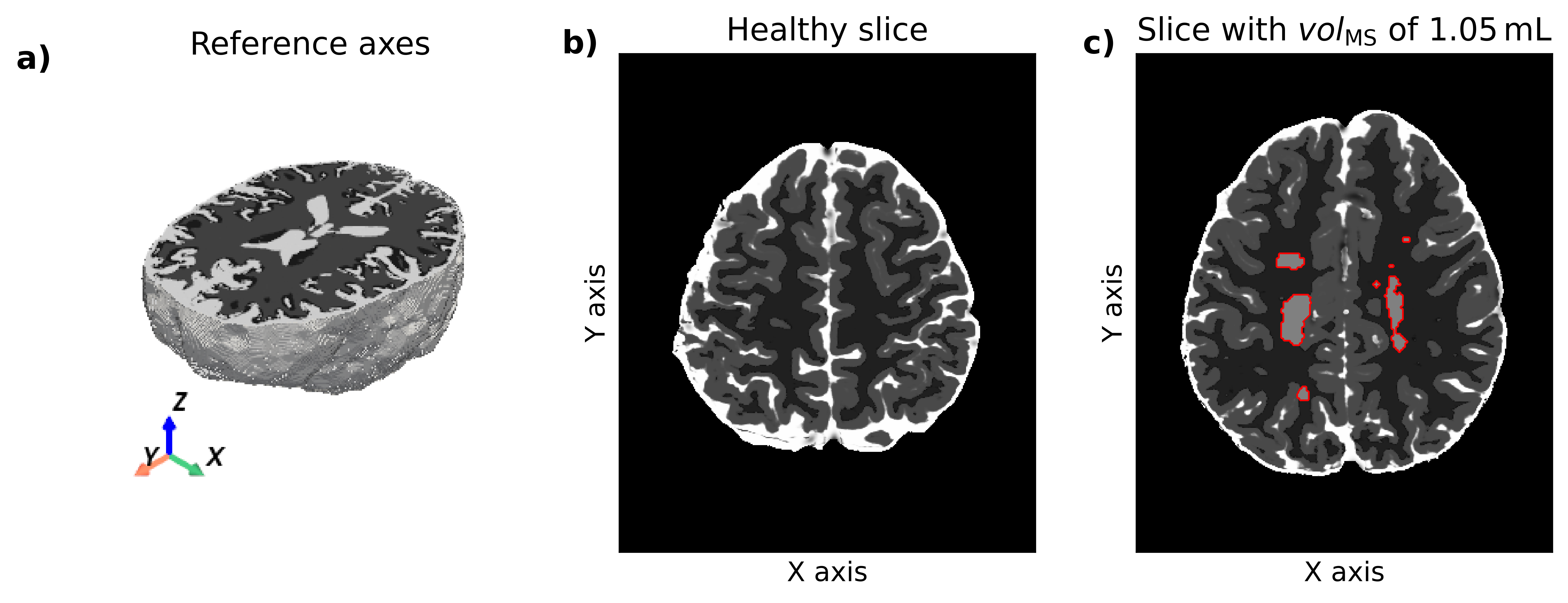}
        \caption{\textbf{Illustration of simulated slices indicating the reference axes. Brighter areas correspond to higher $T_1$ tissues (as CSF), whereas darker tones correspond to low $T_1$ values. The MS lesion is contoured in red.} \textbf{a)} Portion of an original phantom (i.e., before including simulated MS lesions) in the test set, showing the correspondence of axis with the image dimensions. The gradient at a 0$^{\circ}$ spoke was applied along the X-axis, the read-out direction. Slices are stacked along the Z-axis. \textbf{b)} Image of a healthy slice in the test set. The gradient was applied on the X-axis.
        \textbf{c)} Image of a slice with a simulated MS lesion close to the average value ($vol_\text{MS} = 1.05$\,mL).}
        \label{fig:ms_slices}
    \end{figure}

\begin{figure}
        \centering
        \includegraphics[width=\linewidth]{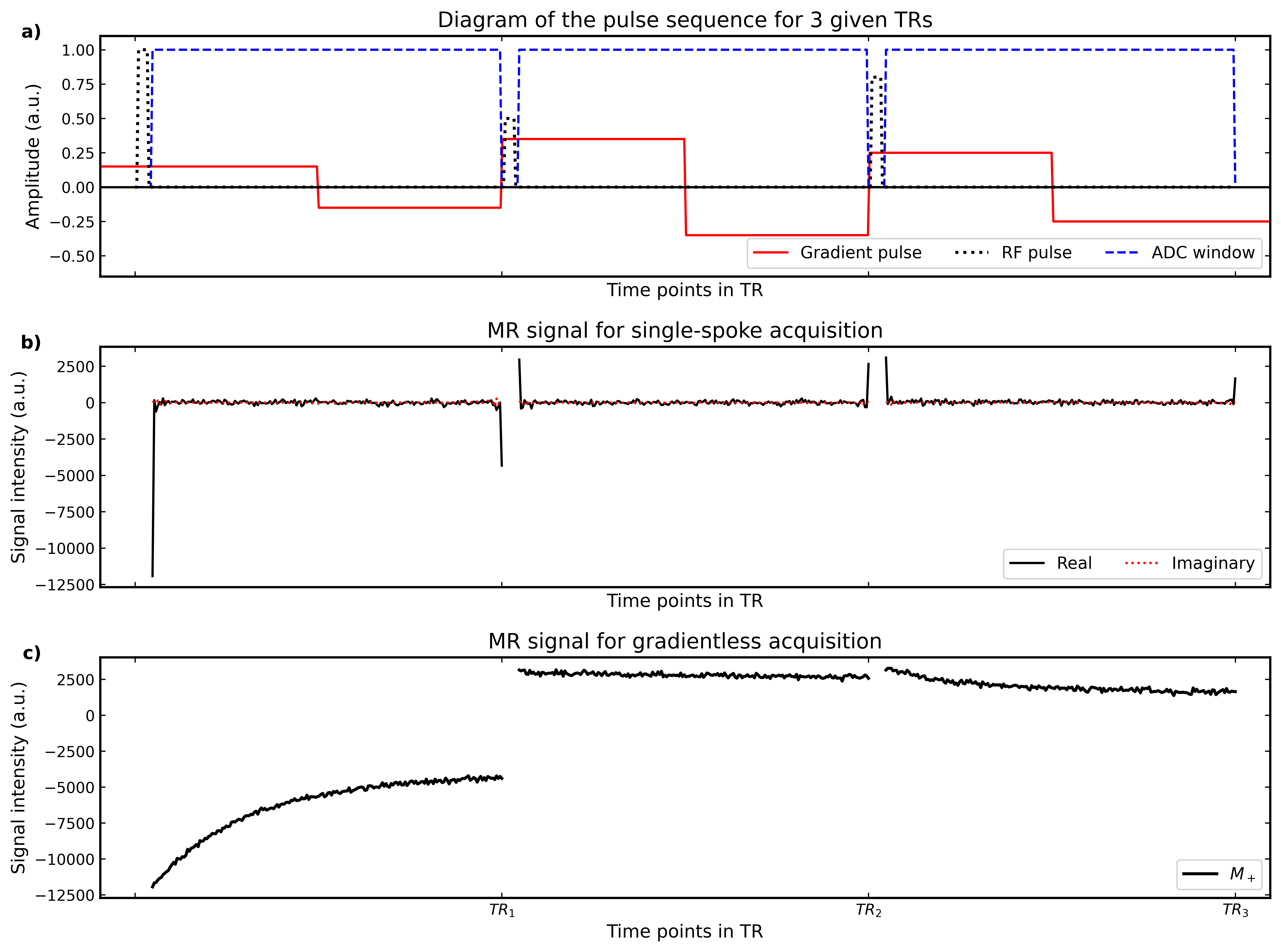}
        \caption{\textbf{Comparison of MR signals in the \emph{single-spoke} and in the \emph{gradientless} setups, showing the first three TRs.} \textbf{a)} Diagram of the MR sequence for three TRs. \textbf{b)} First three TRs of the MR signal for \textit{single-spoke} acquisition with spatial encoding by a single gradient applied in a 0$^{\circ}$ spoke along the X-axis. Solid and dashed lines show, respectively, the real (in-phase) and imaginary (quadrature) components of the complex signal induced by the transverse magnetization decay. \textbf{c)} MR signal for the \textit{gradientless} acquisition, without spatial information, across three TRs. In this case, the FID registered at $B_0$ (with no gradients applied) is observed after each resonant RF pulse, featuring exponential decays that hold the discriminant information between MS and healthy slices. The horizontal axis represents time points at which the global transverse magnetization signal is registered, losing the spatial information encoded in the frequency gradient from \emph{single-spoke}.}
        \label{fig:mrsignals}
    \end{figure}

    Figure~\ref{fig:ms_slices} describes the relation of axes and image dimensions. The read-out axis was X, enabling frequency codification along this direction when the gradient was applied at a spoke of 0$^{\circ}$. 
	Appendix~\ref{app:reprod_sim} contains all the details for reproducing these steps. Simulating 1D MR signals meant following three major steps:
    \begin{enumerate}
        \item Generating a dataset with healthy and MS-affected slices. We used 17 publicly available brain phantoms with healthy tissues, i.e. White Matter (WM), Grey Matter (GM) and Cerebrospinal Fluid (CSF). The information of each brain consisted of four files: three with the amounts of each healthy tissue per voxel (partial volumes allowed), and a fourth one with MS lesion volumes generated by us with a simulation tool \cite{aubert2006twenty}. The final synthetic phantom set contained 935 slices (55 per brain), with nearly 40\,\% containing MS lesions (Supplementary Figure~\ref{suppfig:ms-slices-pie-box}). Treating slices as independent instances provided substantially more samples for training and testing than in a whole-brain approach, which is critical when working with data-driven predictive models.
        \item Optimising the MR signal acquisition parameters, namely inversion-recovery times (TI), repetition times (TR) and flip angles (FA). This step required knowing tissue-specific relaxation times to minimise a cost function prioritising MS signal discrimination while keeping overall tissue distinguishability. Table~\ref{tab:t1t2values} contains $T_1$ and $T_2$ values reported for WM, GM, CSF and MS at 1.5 T \cite{bottomley1984review,larsson1988vivo,o2022vivo}. The simulated acquisitions used a ZTE-like sequence \cite{madio1995ultra}, where a radial $k$-space spoke is encoded after an initial IR pulse. A train of MRF-like RF pulses is then applied, introducing a rewind gradient pulse at half the TR. Consequently, each readout consisted of a spoke travelling from the centre of $k$-space to $k_\text{max}$, followed by a return spoke from $-k_\text{max}$ back to the centre after the refocusing pulse. The MR sequence parameters were optimised using Julia's \texttt{BlackBoxOptim} \cite{Feldt_2023} package, and based on an MRF discrimination (see Supplementary Figure~\ref{suppfig:MRF_opt_N_TR}), setting a schedule of 30 different repetitions at varying TRs.
        \item Simulating the 1D MR signal acquired for each generated slice when the optimised acquisition sequence of pulses (Figure~\ref{fig:mrsignals}a)) was applied. The total scan time of simulated acquisitions was about 3 seconds. 
    \end{enumerate}

	\begin{table}[!h]
		\centering
		\begin{tabular}{cccccc}
        \toprule
			\textbf{Tissue} & \textbf{WM} & \textbf{GM} & \textbf{CSF} & \textbf{MS} \\\midrule
			$T_1$ (ms) & 510 & 1100 & 3831 & 1315 \\
			$T_2$  (ms) & 67 & 77 & 1900 & 174 \\\bottomrule
		\end{tabular}
		\caption{\textbf{Relaxation times values for each tissue, expressed in milliseconds (ms).} WM: White Matter; GM: Grey Matter; CSF: Cerebrospinal Fluid; MS: Multiple Sclerosis. Values reported at 1.5 T in Refs. \cite{bottomley1984review,larsson1988vivo,o2022vivo}}
		\label{tab:t1t2values}
	\end{table}
  
	This sequence of steps was followed twice, with and without a gradient along a single spatial direction for minimal spatial information encoding.  Figure~\ref{fig:mrsignals} shows the MR signals for the acquisition with spatial information (\emph{single-spoke}, top panel) and without (\emph{gradientless}, bottom panel). \emph{Single-spoke} acquisitions provide minimal spatial encoding and produced 1D MR signals that differ notably from the \textit{gradientless} dataset (i.e., without spatial information). In the latter case, the data correspond to FIDs registered for each TR at $B_0$, so the transverse decay is notably slower.

    For each slice, the simulated MR signals were represented as vectors $\vec{x}_{R\cdot T}\in \mathbb{C}$, where $R$ denotes the number of repetitions (TRs) in the MR sequence and $T$ represents the number of points contained in each TRs signal. In the \emph{single-spoke} signal, the $T$ points in each TR contain spatial encoding brought by the gradient. For the \emph{gradientless} data, the spatial encoding is absent, and an FID of the whole tissue under the same $B_0$ field is measured, with the $T$ values corresponding to time points. 
    White noise was added to simulate more realistic acquisition conditions based on the signal-to-noise ratio (SNR) of one of our portable low-field scanners, PhysioMRI Gen I ($B_{0} \approx 72$\,mT, $\text{max}(|\vec{x}|)/\sigma_\text{noise}\approx 500$) \cite{Guallart-Naval2022,guallart2023benchmarking}.

    In addition to these simulations, we explored the imageless framework stability under more realistic or challenging conditions, considering three different scenarios: decreasing the signal-to-noise ratio (SNR), omitting parts of each TR's signal, and considering variability among the $T_1$ and $T_2$ values fed into the data simulation.
	\subsection*{Models for lesion volume estimation}
	\label{sec:method_cnn}
    The imageless framework does not impose any particular model. Depending on the available data and the a priori knowledge, some techniques might better suit specific use cases. Among potential candidates, Convolutional Neural Networks (CNNs) are widely used in various other contexts, excelling at capturing patterns through a cascade of convolutional filters, increasing the pattern complexity with the network's depth \cite{Lecun2015}. In our setting, each voxel is represented by a time signal characterised by tissue-specific spin density, $T_1$ and $T_2$ values that exhibit different temporal patterns if tissue properties change. This change in patterns of exponential decays along the sequence holds diagnostic value, considering that, to obtain the time series, the sequence of MR pulses has been optimised to maximise the distinguishability of our tissue of interest -- MS lesions in this particular case. 
    Stacking several convolutional layers aggregates low-level details (i.e., subtle changes in time signals) into a higher macroscopic level: the volume of MS within the sample in this particular case. For this reason, after extracting relevant features with convolutional layers, these are flattened and combined by fully connected layers (a shallow multilayer perceptron, MLP), enabling their non-linear combinations to produce a final continuous output, in our case, a single-point estimate with the MS lesion volume. This estimated lesion volume is then passed through a threshold ($c_\mathrm{MS}$) to turn the regression outcomes into a binary result, i.e., an MS detection outcome. We used the continuous outcome ($\hat{vol}_\mathrm{MS}$) and the discrete one ($\hat{vol}_\mathrm{MS}>c_\mathrm{MS}$) to compute a combined loss function used to optimize network's weights during training.
    
    We opted for multichannel one-dimensional (1D) CNNs, as the dataset consisted of 200-dimensional vectors representing either the global FID (i.e., \textit{gradientless} MR data) or the FID registered with a single gradient spatial encoding (i.e., \textit{single-spoke} MR data). The 30 repetition times (TRs) were treated as distinct input channels, enabling the CNNs to extract different features for each TR. This is aligned with the MR data acquisition process, since the signal for each TR was obtained by applying a different Flip Angle and at different times, which will result in different shapes for the exponential decays for each TR.
    To accommodate the real and imaginary components of \textit{single-spoke} MR data --the \emph{gradientless} imaginary component is null--, we encoded them as different channels \cite{brooks2019complex}, resulting in $\mathcal{X}_{60\times 200}$ \emph{single-spoke} input tensors and in $\mathcal{X}_{30\times 200}$ \emph{gradientless} input tensors.

    To fit CNNs, the dataset was initially divided into 15 phantoms for training and 2 for testing. Ten cross-validation folds were applied within the training set, leaving one patient out each. The training set was used as well to estimate the parameters for z-score normalization (i.e., subtracting the global mean and scaling by the global standard deviation) before presenting it to the network. This preprocessing step places every feature on a comparable offset and scale, so gradients propagate evenly through the network and improve numerical stability. Cross-validation outcomes for \textit{gradientless} and \textit{single-spoke} data can be found in Supplementary Figures~\ref{fig:cv-gradientless}~and~\ref{fig:cv-singlespoke}, respectively. This process enabled us to explore architectures varying the number of convolutional layers, filters, kernel sizes and other hyperparameters, evaluating the generalisation ability of each setup. Table~\ref{tab:cnn_arch} shows the final architectures for each data set. Supplementary Figure~\ref{fig:arch-singlespoke}~and~\ref{fig:arch-gradientless} illustrate both architectures for \emph{single-spoke} and \emph{gradientless} acquisitions, respectively. Both used ReLU (Rectified Linear Unit) activation functions and a 10\,\% dropout in their dense layers. The chosen architecture was then validated against the two unseen phantoms in the test dataset, double-checking a robust estimate of the network's generalisation performance. 

    \begin{table}[!h]
        \centering
        \begin{tabular}{ccccc}
        \toprule
            \textbf{MR data} & \textbf{Conv. layers} & \textbf{Kernels} & \textbf{Kernel length} & \textbf{Kernel stride}\\\midrule 
            \textit{Single spoke} & 3 & 64, 64, 64 & 3, 3, 3 & 1, 2, 2\\
            \textit{Gradientless} & 2 & 64, 128 & 3, 3  & 1, 2\\\bottomrule
        \end{tabular}
        \caption{\textbf{CNN architectures for the \textit{gradientless} and the \textit{single-spoke} data.} Both architectures presented a single dense layer at the end of the network, connecting the features from the last convolutional layer to the final neuron predicting the MS lesion volume, $\hat{vol}_{\text{MS}}$.}
        \label{tab:cnn_arch}
    \end{table}
    
    Our CNNs infer the MS lesion volume ($\hat{vol}_{\text{MS}}$) and presence in a given slice directly from the input data, without providing any a priori knowledge about the physics governing this relationship. Since this work relies on simulated data, we believe CNNs are robust candidates expected to perform adequately when more realistic data --from real phantoms or even in-vivo measurements-- is available, provided that there is enough data volume. Yet, we compared with two physics-informed models to estimate the presence and MS lesion volume in a given slice: 
    \begin{itemize}
        \item Algebraic Reconstruction Technique (ART, \cite{karczmarz1937angenaherte})) is a general-purpose iterative solver for large systems of linear equations. It was first used in the medical imaging context for X-ray computerised tomography \cite{gordon1970algebraic} and has since been used in a wide range of inverse problems. In the context of our MR data, we cast Bloch equations, which assume the voxel magnetization can be decomposed as a linear combination of exponential functions representing the individual contribution of different types of tissues. This model also assumes that $T_1$ and $T_2$ times governing the exponential decay for the longitudinal ($M_0\left(1-e^{-t/T_1}\right)$) and the transverse ($M_0e^{-t/T_2}$) relaxations, are known (see Table~\ref{tab:t1t2values}). The method iteratively estimates proton densities ($\rho$) in each tissue by minimising the error between the simulated signal and the algorithm output. 
        \item Differential Evolution (DE, \cite{storn1997differential}) also relies on Bloch equations but does not assume known $T_1$ and $T_2$ values. Instead, Black Box optimisation (BBO) searches for an optimum combination of relaxation times and tissue amounts, minimising the total reconstruction error. This search for the optimal $T_1$, $T_2$ and $\rho$ values is performed for each slice, enabling a more realistic scenario in which relaxation times vary across dataset instances, instead of being hard-coded. 
    \end{itemize}

    The key aspect is that, whereas ART and DE utilise a priori knowledge about underlying physics, CNNs rely on convolutional filters to learn how to predict the MS volume (a purely data-driven approach). Embedding explicit physics-based constraints could be especially helpful with real-world measurements, as they narrow the solution space, softening the effect of noise and other uncontrolled artefacts typical of real-world acquisitions. Yet, purely data-driven models might be more flexible and better adapt to the reality represented by the data, reducing the weight of theoretical assumptions that might not be helpful in less ideal scenarios. The performance of these three models for MS lesion volume estimation was compared with simulations including $T_1$ and $T_2$ inter-patient variability in Section~\ref{sec:robustness}.  Their performance comparison without inter-patient variability in relaxation times--when all models are expected to work at their best--can be found in Appendix~\ref{app:test_set_results}. 

	\section{Results}
	\label{sec:results}
	The volume of MS within a given sample, referred to as $vol_\text{MS}$, parameterises our clinical enquiry of interest. Two types of questions can be formulated upon $vol_\text{MS}$: we can quantify the amount of MS --solving a regression problem-- or detect whether MS tissue exists or not --discretise the problem into a classification one based on setting a threshold on the predicted MS volume. For this reason, results are reported with metrics referring to both tasks. On the one hand, regression was evaluated by metrics $R^2$, $b_0$ and $b_1$ report the goodness-of-fit, intercept and slope coefficients, respectively, of the regression model $\hat{vol}_\text{MS} = b_0 + b_1\cdot vol_\text{MS}$, assessing the lesion volume estimation performance. On the other hand, the classification of slices was evaluated by the Area Under the Curve (AUC), the True Positive Rate (TPR, percentage of detected MS slices) and the False Positive Rate (FPR, percentage of misclassified Healthy slices), assessing the lesion detection performance. Lastly, the $vol_\text{MS}^{\text{FN}}$ reports the maximum undetected volume among the False Negatives (FN), giving an idea of the clinical relevance of models' inaccuracies. Besides, to contextualise undetected MS lesions, we also provide the volume of the maximum False Negatives, which represents the biggest in silico MS lesion missed by predictive models. This section reports only on the test-set results, but further information on the cross-validation results can be found in Appendix~\ref{app:crossval}. 
	
    \subsection*{Validation of IMRD for lesion detection and volume estimation}\label{sec:results_single_spoke}
\begin{figure}[!h]
        \centering
        \includegraphics[width=\linewidth]{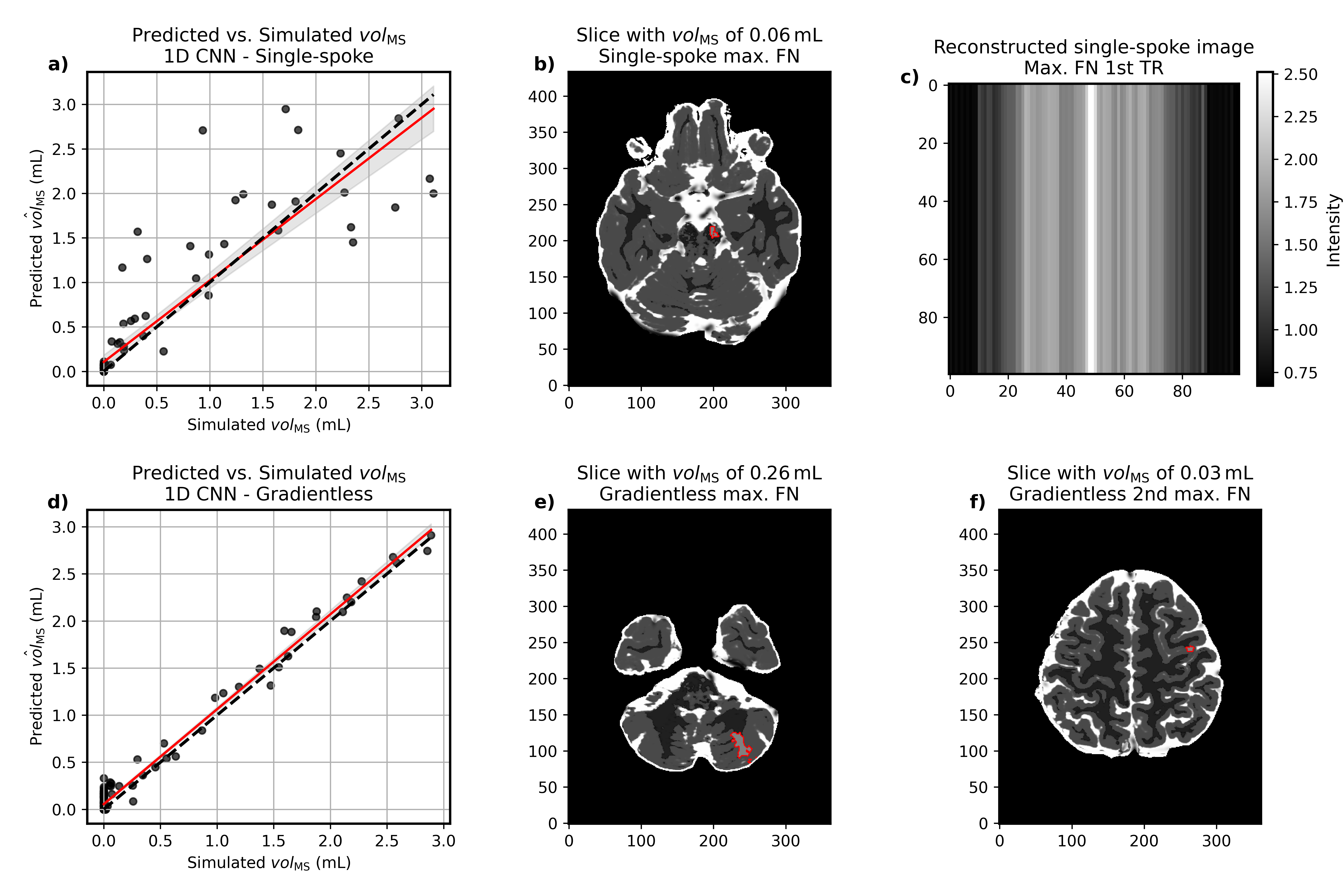}
        \caption{\textbf{Validation of lesion volume estimation for \textit{single-spoke} and \textit{gradientless} data.} \textbf{a)} Regression results for \textit{single-spoke} acquisition, with predicted ($\hat{vol}_{\text{MS}}$) versus simulated lesion volumes ($vol_{\text{MS}}$) showing increased prediction variability for $vol_{\text{MS}}>0$. \textbf{b)} Example of a brain slice with black pixels representing a simulated MS lesion volume $vol_\text{MS}$ of $\approx 0.06$\,mL, the maximum undetected volume (False Negative) for the \textit{single-spoke} model. \textbf{c)} Reconstructed image from first TR for the slice with a simulated MS volume $\approx 0.06$\,mL, which is the maximum undetected volume with the \textit{single-spoke} setup. \textbf{d)} Regression results for a \textit{gradientless} acquisition, with predicted ($\hat{vol}_{\text{MS}}$) versus simulated lesion volumes ($vol_\text{MS}$) showing high agreement with an $R^2$ and a slope close to unity. \textbf{e)} Example of a brain slice with black pixels representing a simulated MS lesion volume $vol_{\text{MS}} \approx 0.26$\,mL, the maximum undetected volume (False Negative) for the \textit{gradientless} model. \textbf{f)} Example of a brain slice with black pixels representing a simulated MS lesion volume $vol_{\text{MS}} \approx 0.03$\,mL, the second maximum undetected volume (False Negative) for the \textit{gradientless} model.} 
        \label{fig:testset-performance}
    \end{figure}

    \emph{Single-spoke} and \emph{gradientless} MR data were separately used to train and test the models for MS volume estimation and detection of MS lesions. Figure~\ref{fig:testset-performance} and Table~\ref{tab:performance_metrics_test} report the performance of CNNs for both acquisition strategies with slices in the test set. Maintaining a single spoke for minimal spatial encoding (upper row in Figure~\ref{fig:testset-performance}) resulted in an $R^2$ of $0.7911$ and an AUC of 0.9536 (Table~\ref{tab:performance_metrics_test}).  
    The biggest undetected MS lesion with \emph{single-spoke} was $vol_\text{MS}^\text{FN} \approx 0.06$\,mL (upper central plot in Figure~\ref{fig:testset-performance}). The \emph{gradientless} acquisition, without any spatial information (low plots in Figure~\ref{fig:testset-performance}), yielded an $R^2$ of 0.985 and an AUC $\approx 0.8$ in terms of detection (Table~\ref{tab:performance_metrics_test}). Despite the maximum undetected lesion volume of $vol_\text{MS}^\text{FN}\approx 0.26$\,mL (lower middle plot in Figure~\ref{fig:testset-performance}), the rest of the false negatives correspond to small lesions below 0.05\,mL (lower right plot in Supplementary Figure~\ref{fig:residuals}).
	
    \begin{table}[!h]
		\centering
		\begin{tabular}{cccccccc}
        \toprule
			\textbf{MR data} & $\mathbf{R^2} \uparrow$ & $\mathbf{b_0}\downarrow$& $\mathbf{b_1}\uparrow$& \textbf{AUC} $\uparrow$ & \textbf{TPR} $\uparrow$& \textbf{FPR}$\downarrow$ & $\mathbf{vol_\text{MS}^\text{FN}}\downarrow$ \\\midrule
                \textit{Single-spoke} & 0.7911 & 0.1435 & 0.9146 & 0.9536 & 0.9211 & 0.0139 & 0.06\\
			\textit{Gradientless} & 0.985 & 0.0797 & 1.0081 & 0.7982
            & 0.825 & 0.1571 & 0.26\\
			\bottomrule
		\end{tabular}
		\caption{\textbf{Model performance metrics in the test set for \textit{gradientless} and the \textit{single-spoke} acquisitions.} Results are obtained against the same two phantoms in the test set. Upside ($\uparrow$) and downside ($\downarrow$) arrows indicate whether if higher (closer to 1) or lower (closer to 0) values, respectively, are better for each metric. The first three metrics ($R^2$, $b_0$ and $b_1$) report the goodness-of-fit, intercept and slope coefficients, respectively, of the regression task. The next three coefficients report the Area Under the Curve (AUC), the True Positive Rate (TPR, percentage of detected MS slices) and the False Positive Rate (FPR, percentage of misclassified Healthy slices), assessing the lesion detection performance. Lastly, the ${vol_{\text{MS}}^{\text{FN}}}$ reports the maximum undetected volume among the False Negatives (FN), giving an idea of the clinical relevance of models' inaccuracies. }
		\label{tab:performance_metrics_test}
    \end{table}
    
	\subsection*{Robustness tests}
    \label{sec:robustness} 
\begin{figure}
        \centering
        \includegraphics[width=\linewidth]{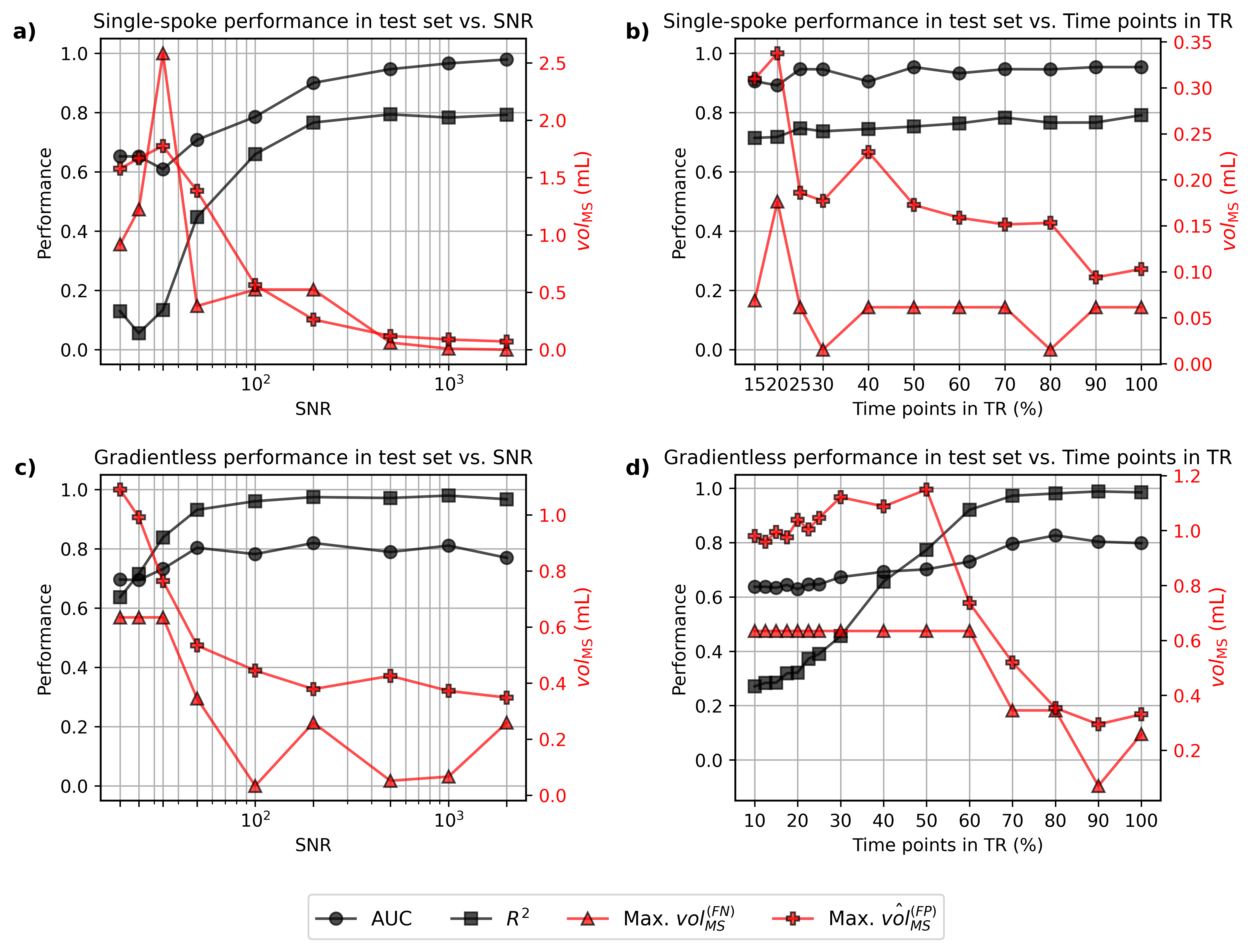}
        
        \caption{\textbf{Performance of CNNs for MS lesion volume estimation and detection under varying SNR and apodization levels.} The left axis scale corresponds to model performance metrics ($AUC$ and $R^2$, black circles and square markers, respectively). The right axis scale corresponds to mL of either simulated MS volume (for max. $vol_\text{MS}^\text{FN}$) or predicted MS volume (for max. $\hat{vol}_\text{MS}^\text{FP}$) \textbf{a)} \emph{Single-spoke} performance metrics as a function of SNR. \textbf{b)} \textit{Single-spoke} performance metrics as a function of the high-frequency information included, which is equivalent to considering regions with a higher applied gradient. \textbf{c)} \textit{Gradientless} performance metrics as a function of SNR. \textbf{d)} \textit{Gradientless} performance metrics as a function of the apodization level applied to the FID signal, losing last instants, when the magnetization vector is already stable.}
        \label{fig:all-snr-apod}
    \end{figure}

    Results within this section report on the performance of different models under less ideal conditions for MR data acquisition. Figure~\ref{fig:all-snr-apod} illustrates the results for SNR (left plots) and information-loss experiments (right plots) for \emph{single-spoke} (upper plots) and \emph{gradientless} (lower plots) MR data acquisitions. Aside from performance metrics ($AUC$ and $R^2$), the maximum undetected lesion volume (max. $vol_\text{MS}^\text{FN}$) and the maximum False Positive estimation (max. $\hat{vol}_\text{MS}^\text{FP}$) were included as well to track the relevance of predictive errors for both MS-affected slices and for healthy slices, respectively. False Positives are healthy slices whose $vol_\text{MS} = 0.0$\,mL but had a $\hat{vol}_\text{MS}>0$ prediction. From the $\hat{vol}_\text{MS}^\text{FP}$ value onwards, all $\hat{vol}_\text{MS} > 0$ truly correspond to slices with MS lesions. 
    
    When the SNR was decreased from ``noise-less'' up to SNR values of 10, \emph{single-spoke} performance (upper left plot in Figure~\ref{fig:all-snr-apod}) drastically dropped to an $AUC\approx0.6$ and a $R^2\approx0.2$, with the latter remaining somewhat stable up to $\text{SNR} = 500$ the reference point used in ``normal condition'' experiments. 
    The \emph{gradientless} performance (lower left plot in Figure~\ref{fig:all-snr-apod}) held stable up to $SNR = 50$ values when both drop to $AUC\approx0.7$ and the $R^2\approx0.6$ values. Both the $vol_\text{MS}^\text{FN}$ and $\hat{vol}_\text{MS}^\text{FP}$ curves responded similarly in both acquisitions, increasing with lower SNR values. The biggest missed MS volumes across the range of SNR values were $2.5$\,mL and $0.6$\,mL for \emph{single-spoke} and \emph{gradientless}, respectively. 
    
    The right plots in Figure~\ref{fig:all-snr-apod} show the results of the apodization experiments. The \emph{single-spoke} performance (upper right plot in Figure~\ref{fig:all-snr-apod}) remains at close to reference $R^2\approx 0.7$ and $AUC\approx 0.9$ values when the highest 85\,\% of frequencies are removed. For the \emph{gradientless} MR data, (lower right plot in Figure~\ref{fig:all-snr-apod}), both the $AUC$ and the $R^2$ increase as the first 60-70\,\% of time points within the TRs are included, reaching $R^2 = 0.98$ values. Beyond these values, the performance remains stable, and the $vol_\text{MS}^\text{FN}$ and the $\hat{vol}_\text{MS}^\text{FP}$ curves show a decay from their maximums (0.6\,mL and 1.2\,mL, respectively), until reaching the reference values achieved when all time points within each TR were considered.

    Finally, inter-slice variability experiments allowed each slice to have distinct $T_1$ and $T_2$ values, drawn from distributions with literature-based mean and variability values (reported in Supplementary Figure~\ref{fig:scatterplot_t1_t2_tissuevar}). Table~\ref{tab:performance_interpat_var} displays the performance metrics for all model candidates with the \emph{gradientless} MR simulated data. The most physics-informed candidate, ART, yields the lowest $R^2$ and $AUC$, of 0.5311 and 0.5023, respectively. The best MS volume estimation and lesion detection were obtained by DE ($R^2$ of 0.8741 and $AUC$ of 0.7388). The regression of 1D CNNs with a combined loss function yielded an $R^2$ of 0.7092 and an AUC of 0.7084. A solely classifying CNN achieved a similar AUC to DE's in MS lesion detection (AUC of 0.7324).

    \begin{table}[!h]
		\centering
		\begin{tabular}{cccccccc}
        \toprule
			\textbf{Model} & $\mathbf{R^2} \uparrow$ & $\mathbf{b_0}\downarrow$& $\mathbf{b_1}\uparrow$& \textbf{AUC} $\uparrow$ & \textbf{TPR} $\uparrow$& \textbf{FPR}$\downarrow$ & $\mathbf{vol_\text{MS}^\text{FN}}\downarrow$\\\midrule
			\textit{ART} & 0.5311 & 0.5529 & 0.9991 & 0.5023 & 0.9756 & 0.9710 & 0\\
                \textit{DE} & 0.8741 & 0.0917 & 0.9396 & 0.7388 & 0.9268 & 0.4493 & 0.08\\
                \textit{1D CNN (regr.)} & 0.7092 & 0.1387 & 0.721 & 0.7084 &0.6341 &  0.2174 & 0.54\\
                \textit{1D CNN (det.)} & -- & -- & -- & 0.7324 & 0.6098 & 0.1449 & 0.54\\
			\bottomrule
		\end{tabular}
		\caption{\textbf{Model performance metrics in the test set for \textit{gradientless} acquisition with inter-slice variability in sampled relaxation times.} Results are obtained against the same two phantoms in the test set. Upside ($\uparrow$) and downside ($\downarrow$) arrows indicate whether if higher (closer to 1) or lower (closer to 0) values, respectively, are better for each metric. See details of ART and DE performance without variability in Table~\ref{tab:performance_metrics_test}, and information about $T_1$ and $T_2$ distributions in the caption of Supplementary Figure~\ref{fig:scatterplot_t1_t2_tissuevar}.}
		\label{tab:performance_interpat_var}
    \end{table}
    
	\section{Discussion}
	\label{sec:dicuss}
    This work aims to introduce a framework for imageless MR diagnosis for efficient and cost-effective decision-making in clinical contexts. Along with the concept motivation and definition in the first sections, we included an in-silico toy example with simulated MS lesions, hoping to provide a potential illustration of our imageless proposal in practice. Besides, we contemplated two virtually viable imageless MR setups that could be implemented for such tasks, highlighting that the imageless framework encapsulates not only different clinical questions (see Section~\ref{sec:intro}) but also allow for different implementations and techniques under the shared feature of substantially reducing hardware requirements imposed by traditional MR imaging.

    Results with simulated MR signals from in-silico MS-affected brains (Figure~\ref{fig:testset-performance}, and Tables~\ref{tab:performance_metrics_test}~and~\ref{tab:performance_art_de}), suggest that including minimal spatial encoding as the \emph{single-spoke} acquisition does, could aid sensitive and precise MS lesion detection ($AUC\approx0.95$ and $FPR\approx0.01$). Yet, \emph{single-spoke} regression performance ($R^2\approx0.8$) was less accurate than for the \emph{gradientless} setup ($R^2\approx0.99$). This suggests that the gradient applied to include spatial information might allow for better event detection at the cost of losing the total contribution of MS tissue within the slice to the signal's magnitude. This information in the \emph{gradientless} signal, substantially improves the accuracy in MS lesion volume estimation. The lower left plot in Figure~\ref{fig:testset-performance} shows how simulated MS volumes ($vol_\text{MS}$) and the predictions ($\hat{vol}_\text{MS}$) fall closely aligned to the perfect prediction diagonal (black dashed line). This coincides as well with the results obtained with ART and DE (Supplementary Figure~\ref{supfig:zerospoke-art-bbo} and Supplementary Table~\ref{tab:performance_art_de}), suggesting that \emph{gradientless} MR signals can be a useful source of information following the imageless principle. As a counterpart, \emph{gradientless} exhibits a higher FPR, visible as well in the over-estimation of cases at $vol_\text{MS}=0$, many of which lie above the dashed diagonal in the lower left plot from Figure~\ref{fig:testset-performance}. This overestimation of some healthy slices seems to be present as well in ART and DE results (Supplementary Figure~\ref{supfig:zerospoke-art-bbo}), which discards its roots in suboptimal CNNs' architecture or hyperparameter tuning. A closer look at healthy slices in Supplementary Figure~\ref{fig:zerospoke-healthyslices} depicts a correlation pattern between MS volume estimations above the classification threshold and the average MR signal intensity, with the latter being an indicator of the overall tissue within a given slice. Hence, \emph{gradientless} MR signals of slices with small MS lesions might overlap with \emph{gradientless} MR signals of healthy slices with more tissue. This would explain why models diagnose ``big'' healthy slices as small MS lesions. Nonetheless, aside from the maximum undetected lesion volume of 0.26\,mL, which seems somewhat outlying--see right plot in Supplementary Figure~\ref{fig:residuals} and Supplementary Figure~\ref{fig:class-gradientless}--, the rest of missed MS-affected slices contained less than 0.05\,mL of affected tissue, highlighting \emph{gradientless} sensitivity.

    The comparison between \emph{single-spoke} and \emph{gradientless} in less stable scenarios also shed light to differences between them. Decreasing the SNR emulates a quite common phenomena in the context of LF MR setups. In these experiments (left plots in Figure~\ref{fig:all-snr-apod}), the \emph{gradientless} acquisition retained higher $R^2$ across decreasing SNR values than the \emph{single-spoke} acquisition. In contrast, \emph{single-spoke} seemed more robust against information loss (right plots in Figure~\ref{fig:all-snr-apod}). It is important to mention that information-loss implied different things for each setup. In \emph{single-spoke}, information in the highest frequencies was removed, meaning that ``detail'' information from the 1D MR signal was removed. On the contrary, for \emph{gradientless} signals, time points at the tail of each TR were removed, which is equivalent to shortening the acquisition windows. Thus, if the information about MS lesions is somewhat localized, one would expect it to be contained in low-frequency components of the \emph{single-spoke} signal, and even if abundant high-frequency details are removed (up to 60-70\,\% of top highest frequencies), one would expect a decent performance as long as low-frequency information is kept on the signal. 
    
    However, in the case of \emph{gradientless} signal, the relevant information spreads across the whole exponential decay of the FID in each TR. Hence, performance will not suffer as long as TR segments where all tissues recovered their longitudinal magnetization, are removed. Otherwise, as soon as time points where the exponential decay is still happening are removed, one could expect a decay in performance. Indeed, this is what the \emph{gradientless} information loss results suggest (lower right plot in Figure~\ref{fig:all-snr-apod}). Including only initial time points allows for a shallow predictive power, probably because these initial moments of each FID carry out information about the overall tissue size. However, the performance (reflected by both the $AUC$ and the $R^2$) clearly improves as time points located along the exponential decay are included, and stabilises in latter time points, when the longitudinal magnetization has likely been recovered. The fact that MS volume is estimated by relaxometry—i.e. from exponential decays governed by $T_2^*$—also explains the lower $R^2$ obtained by \emph{single-spoke} readout. The gradient used in the \emph{single-spoke} acquisition, substantially reducing the transverse magnetization signal, thereby diminishes the relaxation contrast that drives MS-volume predictions.

    Finally, on inter-slice $T_1$ and $T_2$ variability, only the \emph{gradientless} signal was considered because it had shown the best regression performance in ``normal'' conditions, and would therefore allow us to better assess the drop in performance due to the inclusion of relaxation times variability. Results of increasing slices heterogeneity suggest that physics-informed candidates for imageless frameworks should include steps for finding unknown relaxation times, as DE does (Table~\ref{tab:performance_interpat_var}). Accounting for this variability in relaxation times resulted in a more balanced performance of DE ($R^2\approx0.87$) than ART's fixed-$T_1$ and $T_2$ assumptions ($R^2\approx0.53$) and than purely data-driven CNN's performance ($R^2\approx0.71$). A solely classifying 1D CNN was trained as well ($AUC\approx0.73$), but the performance shows a clear drop in sensitivity ($TPR\approx0.61$) compared to reference results without variability ($TPR\approx0.83$, Table~\ref{tab:performance_metrics_test}). However, it is important to mention that CNNs still retain a certain predictive power, and considering the increase in variability was not coupled with any sort of data augmentation, there might be room for CNNs improvement in performance provided that bigger sample sizes are used for training. 

    These results, while promising, also pointed out weaknesses within this particular case study. These can be covered as well in future work, tackling the translation from in-silico to ex-vivo or even in-vivo measurements. For instance, information-loss results with \emph{gradientless} (lower right plot in Figure~\ref{fig:all-snr-apod}) could include refining 1D CNNs via attention mechanisms that selectively weight certain time points, or more radically by performing some time-wise variable selection. Focusing on mid-TR time points might prevent the tissue size effect seen in \textit{gradientless}, balancing to keep the informative relationship between signal amplitude and lesion size, while reducing false positives. 

    In terms of facing relaxation times variability, there might be different lines of work. An obvious one is to enrich the training subset with common post-acquisition techniques (such as data augmentation). This could be applied to improve data-driven models' robustness against variability in the tissue's relaxation times. Furthermore, leveraging data-driven estimates with information about MR's underlying physics seemed to bring more stable outcomes, as seen with DE. Thus, integrating a priori information about MR physics within 1D CNNs could be another path for better models to process imageless data. Nonetheless, aside from model improvements, a potential source of limitation might stem from the information provided to the MR sequence optimisation, which was optimised for single-point $T_1$ and $T_2$ values per tissue. However, real tissues may show stronger overlap in their relaxation times or deviate from average values, especially in pathological tissue. Further investigation on different acquisition strategies with MRF-like sequences accounting for possible $T_1$ and $T_2$ inhomogeneities could also play a favourable role for imageless approaches. Exploring the option of merging information from several spokes in an imageless fashion could enrich the spatial information considered for the MR acquisition sequence optimisation. Besides, the sequence parameters, such as the number of TRs, was optimised based on an MRF discrimination (see Supplementary Figure~\ref{suppfig:MRF_opt_N_TR}), setting the MR sequence accordingly. However, this approach might be suboptimal compared to an MR sequence optimisation that explicitly includes the performance of the final classification models. This would also allow considering each model's performance as the cost function, instead of using the cross-correlation between tissue signals, as we did. Such MR sequences customised for each supervised model, accounting for their discriminative patterns, might improve the outcomes.
    
    As a final note, it is essential to emphasize that this work is an early-stage demonstration of an IMRD framework, still far from a ready-to-implement PoC solution. Simulations with in-silico MS lesions aimed to illustrate the potential of imageless frameworks to reduce hardware requirements, but real-world MR data, collected under more realistic and heterogeneous conditions, will present additional challenges remaining to be tested. Beyond potential improvements of in-silico results, 
    models leveraging several slices as in whole 3D brains, incorporating real $T_1$ and $T_2$ inter-subject variability, validating IMRD with real-world datasets or the inclusion of inhomogeneous zones in our FoV, similar to the ones found in low-cost, portable systems \cite{guallart2023benchmarking, galve2024elliptical}, remain to be tackled in future work. Furthermore, all these aspects might impact differently in other tissue lesions than MS. Hence, exploring more tissue lesions and using real-life data are key steps for future IMRD development and assessing how these factors impact performance. These developments will be key for ensuring the translation of IMRD as a clinical application, including triage and rapid screening scenarios. 
 
\section{Outlook}
\label{sec:outlook}
The proposed IMRD framework questions the need for conventional MR systems, designed to generate high-quality images interpreted by human sight, to answer all sorts of clinical enquiries. This implies reframing MR raw signals not as an intermediate step for final image reconstruction, but as a piece of information holding intrinsic diagnostic value. 

If low-dimensional MR signals, which are not intended to build an image, can answer clinical questions of interest -- as our work suggests -- classical MR systems, governed by hardware constraints imposed by image-centred analysis, might no longer be essential in some cases. The viability of IMRD would then break the monogamy between MR and image visualisation, enabling its use as a direct question-answering modality. Some of the questions that could fit in an IMRD framework are described in Section~\ref{sec:imrd_applications}: detecting the presence of a substance, quantifying the abundance of a tissue or seeking anomalies in the structural disposition of tissues.

In this work, undertaking in-silico MS lesions as a case study, we aimed to illustrate how different elements of an IMRD setup could articulate to answer a clinical question. These elements encompass: a simple hardware--where a single gradient at 0$^{\circ}$ or not even applying gradients--; an algorithm for optimising fast MR pulse sequences using raw MR signals as data representation instead of images; and a model to process the MR data and finally answer the question of interest. Yet, the specific elements chosen in each application might vary, meaning that optimal IMRD setups are likely to be task-specific. In fact, in the simulation MS case study, two potential imageless paths seemed viable--\emph{single-spoke} and \emph{gradientless}. 

Considering the limitations of the work's early stage, the relevant message is that there seem to be feasible IMRD implementations that might support triage, rapid screening or monitoring when a full image reconstruction is not necessary. The implications of a cost-effective and efficient MR usage in clinical scenarios would be game-changing, and that is the potential value of the IMRD notion proposed and illustrated in this work. Future efforts include exploring more conditions--not merely MS--and types of diagnostic questions--detection, grading and/or regression. Yet, as mentioned, future IMRD implementations will probably follow a task-driven design, not only in terms of the final models, but also in terms of MR signal acquisition and requiring an optimal interplay between hardware, physics and AI models for the success of different applications.

    \section{Conclusions}
	\label{sec:conclusion}

Imageless MR Diagnostics (IMRD) offers the potential to significantly reduce hardware complexity and scan times while maintaining diagnostic precision. By bypassing the need for image reconstruction, IMRD challenges traditional solutions in MR-based diagnostics and opens new possibilities for clinical and point-of-care (POC) applications in scenarios where conventional scanners are inaccessible. 

Our results support that an IMRD framework can successfully answer a clinical enquiry--in this case, detecting/quantifying the amount of MS tissue--, even when spatial encoding is heavily reduced (\textit{single-spoke}, achieving an $AUC\approx0.95$) or directly absent (\textit{gradientless}, achieving an $R^2\approx0.98$) and no image can be reconstructed. The suggested viability of relying on one-dimensional signal evolutions alone could be particularly relevant for low-cost portable MR systems without robust imaging components and requirements. Stability experiments with decreasing SNR and strong information loss (apodization) further suggested that imageless MR data can tolerate such hardware and acquisition limitations, supporting potential utility in resource-limited environments. Future work on the particular case of MS could benefit from exploring the potential overestimation of tissue-rich healthy slices (Supplementary Figure~\ref{fig:zerospoke-healthyslices}), present as well in ART's and DE's predictions (Supplementary Figure~\ref{fig:zerospoke-art-bbo}). Yet, robustness tests also pointed out that facing real-world data--with noisier signals and relaxation times variability--constitutes the biggest challenge. Incorporating attention mechanisms, focusing on relevant signal segments, adding physics-informed constraints within models (as DE does) or increasing training sample size with data augmentation could robustify model performance with real-world data.

However, the most important workforce remains to be put on expanding the IMRD framework to answer more clinical questions. This will likely be a task-specific procedure, requiring a different combination of hardware requirements, MR sequence optimisation and data processing models depending on the clinical question tackled in each application. As shown in the MS case, there might be more than one possible IMRD path (e.g.: \emph{single-spoke} or \emph{gradientless} acquisitions, physics-informed or data-driven models) to answer the same clinical question, which illustrates the new range of possibilities that working in an imageless fashion might open, with the substantial reduction in MR-associated costs and constraints as the common factor shared by them all.

In conclusion, this work provides preliminary but compelling evidence of the feasibility of a truly Imageless MR Diagnostic tool that exports MR as a wide diagnostic tool. By shifting towards a direct signal analysis, IMRD frameworks hold promise for simpler hardware designs and near real-time acquisitions, typical of low-field MRI setups. Yet, it is vital to test the approach with clinical enquiries of different nature, knowing beforehand that each casuistic will probably require refining the acquisition sequence design, incorporating domain knowledge if possible, and extensively validating with real-world data. If feasible solutions for each step in the IMRD framework can be found to answer clinical closed questions, as the promising outcomes of our in-silico study suggest, IMRD could take a significant step forward, bringing faster and more accessible MR-based diagnosis to patients globally.

\section{Acknowledgment}
This work was supported by the European Innovation Council under grant 101136407.

\section{Author contributions}
Project conceived by FG, AW, and JA. Simulations by PGC and FG. AI models by AGC, PGC, EI, VVDV, and MS. Data analysis by AGC, PGC, and FG. Writing by AGC, PGC, FG, and JA, with input from all authors.

\section{Conflict of interest}
FG, AW, and JA have a patent on IMRD.
	
    
    \bibliographystyle{ieeetr}
    \bibliography{Bibliografia}

    \newpage
    \appendix
    \renewcommand{\thesubsection}{S\arabic{subsection}}   
    \section{Supplementary material}  

    This section includes further information about technical details and results mentioned across the main text.

    \subsection{Reproducibility of simulations} 
    \label{app:reprod_sim}
    This section gives a more in-depth view of each step in the data simulation process.
    
    \begin{enumerate}
        \item \textbf{Phantom dataset generation:} from the starting set of healthy tissues until obtaining the final data fed to the models, we followed these steps: 
        \begin{enumerate}
            \item Building a set of healthy phantoms. As stated in Section~\ref{sec:method}, our phantoms consisted of 17 brain volumes of matrix size $362\times434\times362$ pixels, storing each healthy tissue (white matter, gray matter and cerebrospinal fluid) in a different file. Given a 30\,cm Field of View (FoV) in the X-axis, and considering voxels were set to be isotropic, having the same size along all dimensions, each voxel corresponded to $vol_\text{voxel} = \left(FOV_X/N_X\right)^3 = (30/362)^3 = 0.082873^3 = 0.000569$ mL.
            \item Building a set with MS phantoms. A fourth file with simulated MS lesions \cite{da2019multiple}, was generated for each brain, resulting in 68 files. 
            \item Depuration of slices. Each brain (represented by four tissues) originally contained 362 slices, but in the end, 55 equidistant slices from each brain were selected. This was done to ensure slices too similar were not over-represented, and to exclude those with scarce or no tissue.  
            \item Obtention of the final phantoms dataset. At this point, there were two subsets of 935 slices: one with healthy brains and another with MS brains. To obtain the final dataset, a Bernoulli distribution (with $p =0.5$) determined whether the MS version of each slice would be included, setting independence between consecutive slices of presenting their healthy or their MS lesion version. The final distribution of MS slices is shown in Supplementary Figure~\ref{suppfig:ms-slices-pie-box}. It is important to mention that in MS phantoms, not all slices presented MS tissue, which explains the final under-representation of MS slices despite setting a 50\,\% chance of sampling from the MS phantoms.
        \end{enumerate} 
        \begin{suppfigure}
        \centering
            \includegraphics[width = \textwidth]{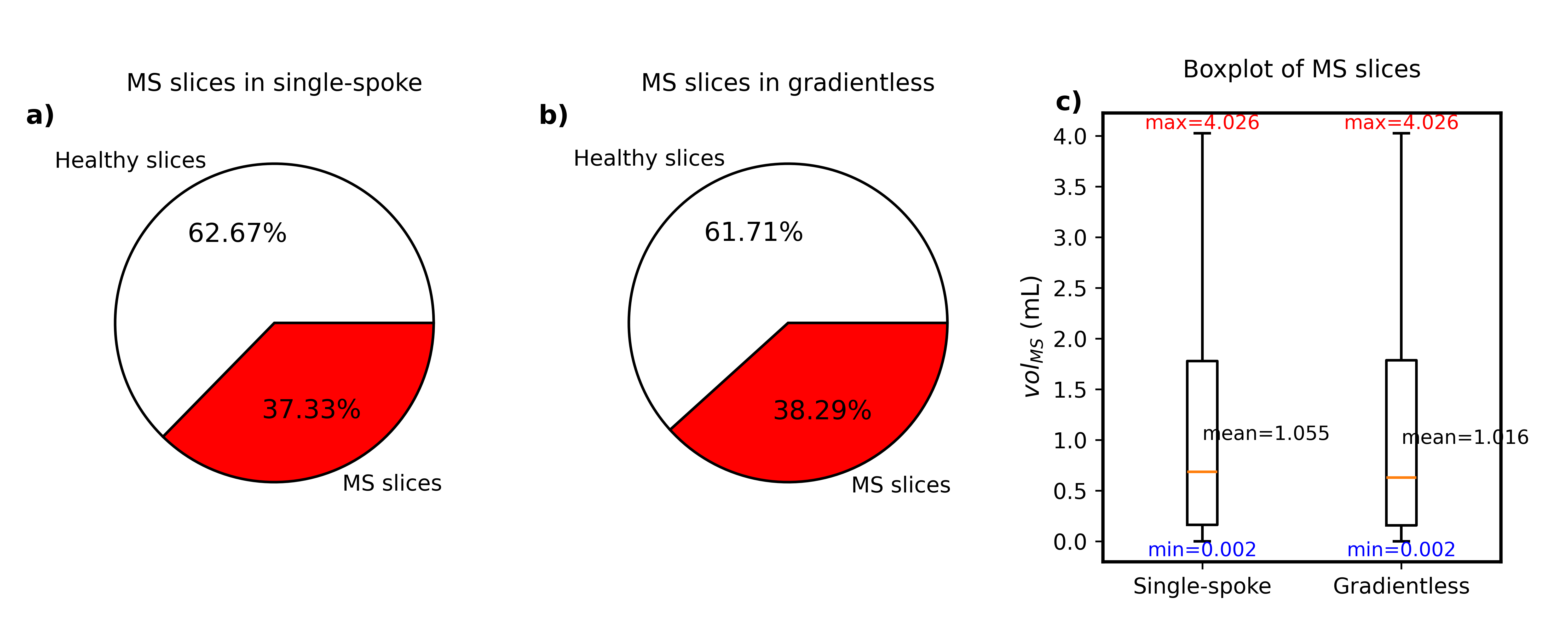}
            \caption{Pie charts (left and center) comparing the proportion of healthy (white) and MS (red) slices in the gradientless (left) and single-spoke (center) datasets. The boxplot on the right depicts the distribution of MS lesion volumes ($vol_\text{MS}$ in mL) for each acquisition setup, with annotated minimum, mean and maximum values.}\label{suppfig:ms-slices-pie-box}
        \end{suppfigure}
        \item \textbf{Sequence design and parameter optimisation.} After obtaining the set of phantoms, their corresponding MR signals could be simulated. 
       First, the setup for the MR sequence had to be optimised. A ZTE sequence was considered, optimising (i) the Inversion Recovery Time, (ii) the Repetition Times (TR), and (iii) the Flip Angles (FA). Once a given radial spoke is set, a succession of RF excitations with different flip angles follows an initial Inversion Recovery pulse. Between excitations, an acquisition window comprising the length of the TR measures the displacement from the signal space center to its maximum in half of the TR. When reached, a rewind gradient is set to return from said maximum to the center. The gradient amplitude ($g_r$) is adjusted for each TR$_r$, so its product equates 4\,ms$\times$1\,mT/m. The \textit{gradientless} setup applied no gradient, as seen in Figure~\ref{fig:mrsignals}.

Parameters were optimised with the default differential evolution algorithm in \texttt{BlackBoxOptim.jl} package, using the Julia programming language, providing as inputs: (i) the tissue's $T_1$ and $T_2$ values; (ii) the length of the RF pulse train; and (iii) the upper and lower bounds of the parameters ($t_\text{IR}\in [0.01,2]$\,s, $\text{TR}\in[10,500]$\,ms and $\text{FA}\in[10,150]^{\circ}$). We use the following cost function:
\begin{equation}
C =\sum_{i=1}^{N_T} \sum_{j=1}^{N_T} |A_{ij} -A_{ii}|+ \frac{1}{N_T} \sum_{i=1}^{N_T} |A_{i,\text{MS}}|- |A_\text{MS,MS}|
\label{eq:cost_function}
\end{equation}
with $N_T$ the amount of tissues considered, namely four (white matter, grey matter, cerebrospinal fluid and multiple sclerosis), and $A_{i,j}=\vec{s}_i\cdot \vec{s}_j /|\vec{s}_i||\vec{s}_j|$ is the cross-correlation matrix \cite{COHEN201715}. The term $\vec{s}_i$ consists of the transverse magnetization $M_t^\text{F}$ at the end of each TR in the train for tissue $i$, following the spirit of MR-fingerprinting. The first term is the signal distinguishability among each pair of tissues, while the second term adds extra distinguishability for MS.

As mentioned in the main text, we chose 30 TRs since we observed good MS discrimination with an MR-fingerprinting procedure, i.e. image-based, up to a combination of 30 TRs and 40 spokes. In MR-fingerprinting, optimising the sequence according to the magnetization values near $k_\text{max}$ is meaningful since each TR is used to produce an image. However, in the IMRD paradigm MS discrimination is signal-based, i.e. it uses all time points in each TR. In preliminary calculations, we have observed that minimization of Eq.~(\ref{eq:cost_function}), with $\vec{s}_i$ containing all timepoints, resulted in a threshold around 15 TRs, with no appreciable improvement beyond that TR length for the gradientless case. Note that if matrix A, with all time points, is nearly diagonal, then each tissue signal is almost orthogonal to each other, which in principle leads to perfect distinguishability, at least with respect to ART and DE algorithms.

\item \textbf{Simulation of the dataset with MR 1D signals:} The generated phantoms and the MR sequence parameters file were then fed into a script simulating the evolution of each slice during the sequence. This script lets the user apply a gradient in a given spatial direction for the \textit{single-spoke} acquisition. In this work, and for the sake of simplicity, a radial angle of $\theta=0^{\circ}$ was set. Each acquisition comprised 200 points per TR, resulting in 6,000 total sampled points per dataset entry for a 30 TR schedule. Since signal simulation is GPU-based, we minimized vRAM use by assigning a single tissue for each pixel with the following criterion: if there is MS in that pixel, we assign $\rho_\text{MS}$ to that pixel, while for the rest of pixels we assign the density of the tissue with the highest density for that pixel.

A 30 TR length was selected using an MRF framework, optimising different parameter sets of varying lengths for use in a radial MRF acquisition, while also evaluating the effect of different radial undersampling levels. The test aimed to determine the percentage of the original MS phantom correctly identified by the MRF algorithm. As shown in Supplementary Figure~\ref{suppfig:MRF_opt_N_TR}, the best trade-off between accuracy and scan time was achieved with a 30 TR sequence. 

        \begin{suppfigure}
            \includegraphics[width=\textwidth]{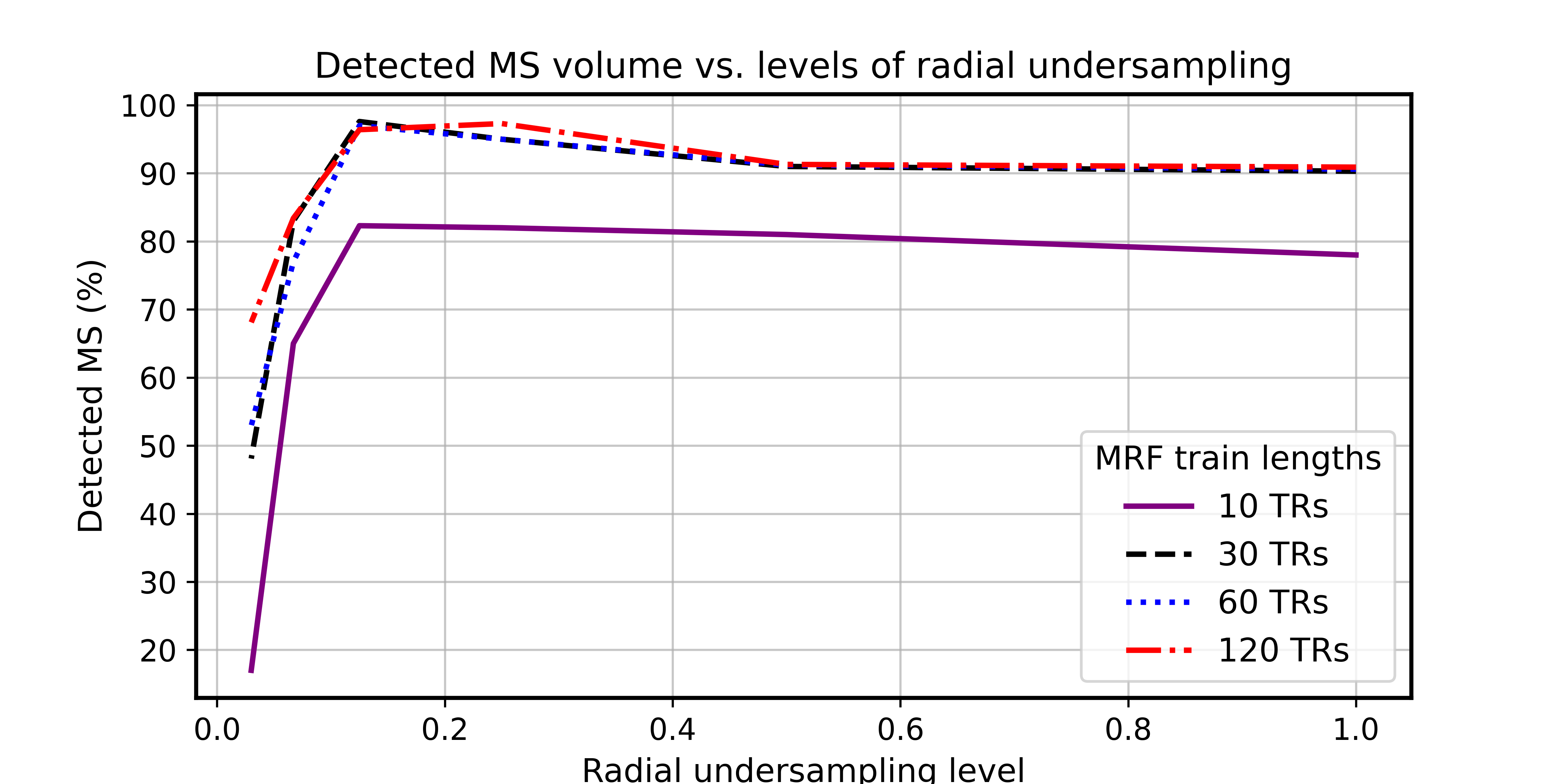}
            \caption{\textbf{Discrimination performance with MRF vs levels of radial undersampling by number of TRs.} The discrimination performance is expressed as the percentage of detected MS pixels for each configuration coupling a number of TRs and undersampling levels. The plot shows a substantial improvement in detected MS from 10 TR to 30 TR acquisitions, with additional TRs yielding a similar performance, suggesting that 30 TR sequences present the best trade-off between acquisition duration and detection performance. }
            \label{suppfig:MRF_opt_N_TR}
        \end{suppfigure}
        
    \end{enumerate}
    
    \subsection{Cross-validation results}
    \label{app:crossval}
    Cross-validation was implemented to assess CNN generalisation with different architectures and hyperparameters. To do so, the model loss curves for training and validation subsets in each CV fold and some regression and classification metrics were assessed. 
    
    This information for the \textit{single-spoke} acquisition is reported graphically in Supplementary Figure~\ref{fig:cv-singlespoke}. The left panels show the model's stable convergence during training, as evidenced by the decline in the combined loss function over the 1,000 epochs. The middle and right panels show metrics reporting the lesion volume estimation and detection performances, respectively. Analogously, Figure~\ref{fig:cv-gradientless} reports the same information but for the \textit{gradientless} case.

    \begin{suppfigure}[H]
        \centering
        \includegraphics[width=\linewidth]{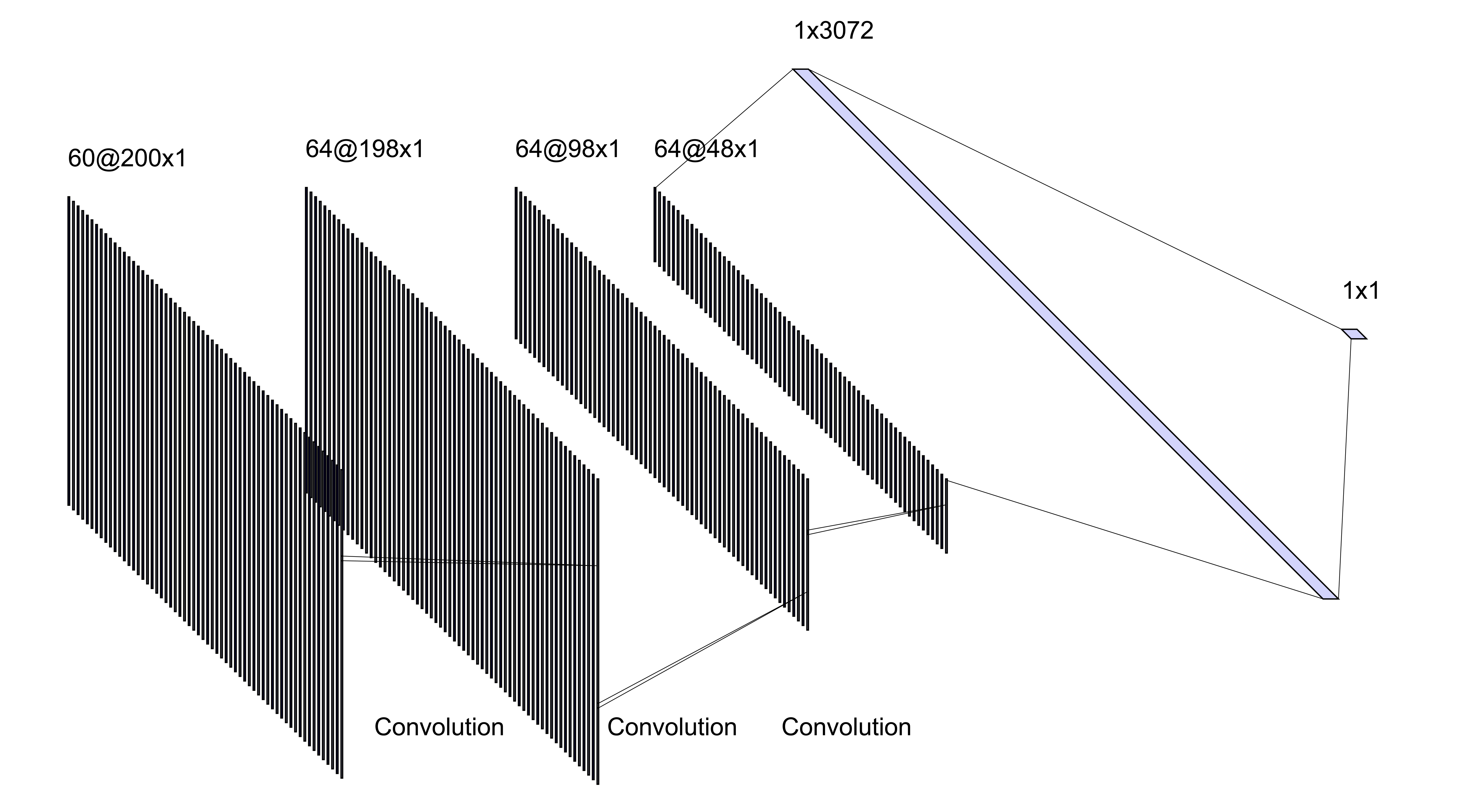}
        \caption{\textbf{1D CNN architecture for \textit{single-spoke} data.}}
        \label{fig:arch-singlespoke}
    \end{suppfigure}
    
    \begin{suppfigure}[H]
        \centering
        \includegraphics[width=\linewidth]{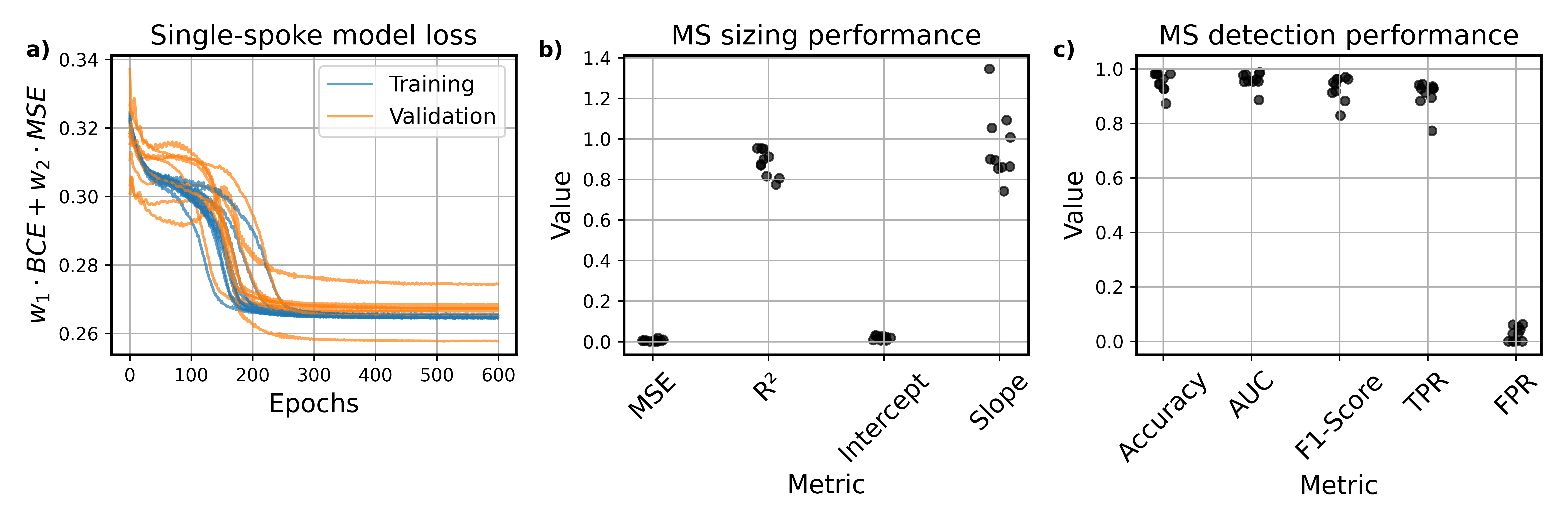}
        \caption{\textbf{Cross-validation results for \textit{single-spoke} data.}\textbf{a) }Model loss during training and validation for 10-fold cross-validation. The consistent convergence indicates robust optimisation and generalisation across folds. \textbf{b) }Performance metrics for lesion volume estimation, showing the values across folds of the mean squared error (MSE), the goodness-of-fit coefficient $R^2$, intercept and slope values of the regression model between the simulated and the predicted MS volume. The metrics distribution shows a less accurate volume prediction than \textit{gradientless} IMRD. \textbf{c) }Performance metrics for the lesion detection, including accuracy, AUC, F1-score, TPR, and FPR. Despite an outlying fold yielding lower performance metrics, the lesion detection task is less affected than the volume estimation, with a performance similar to \textit{gradientless} lesion detection.}
        \label{fig:cv-singlespoke}
    \end{suppfigure}

    \begin{suppfigure}[H]
        \centering
        \includegraphics[width=\linewidth]{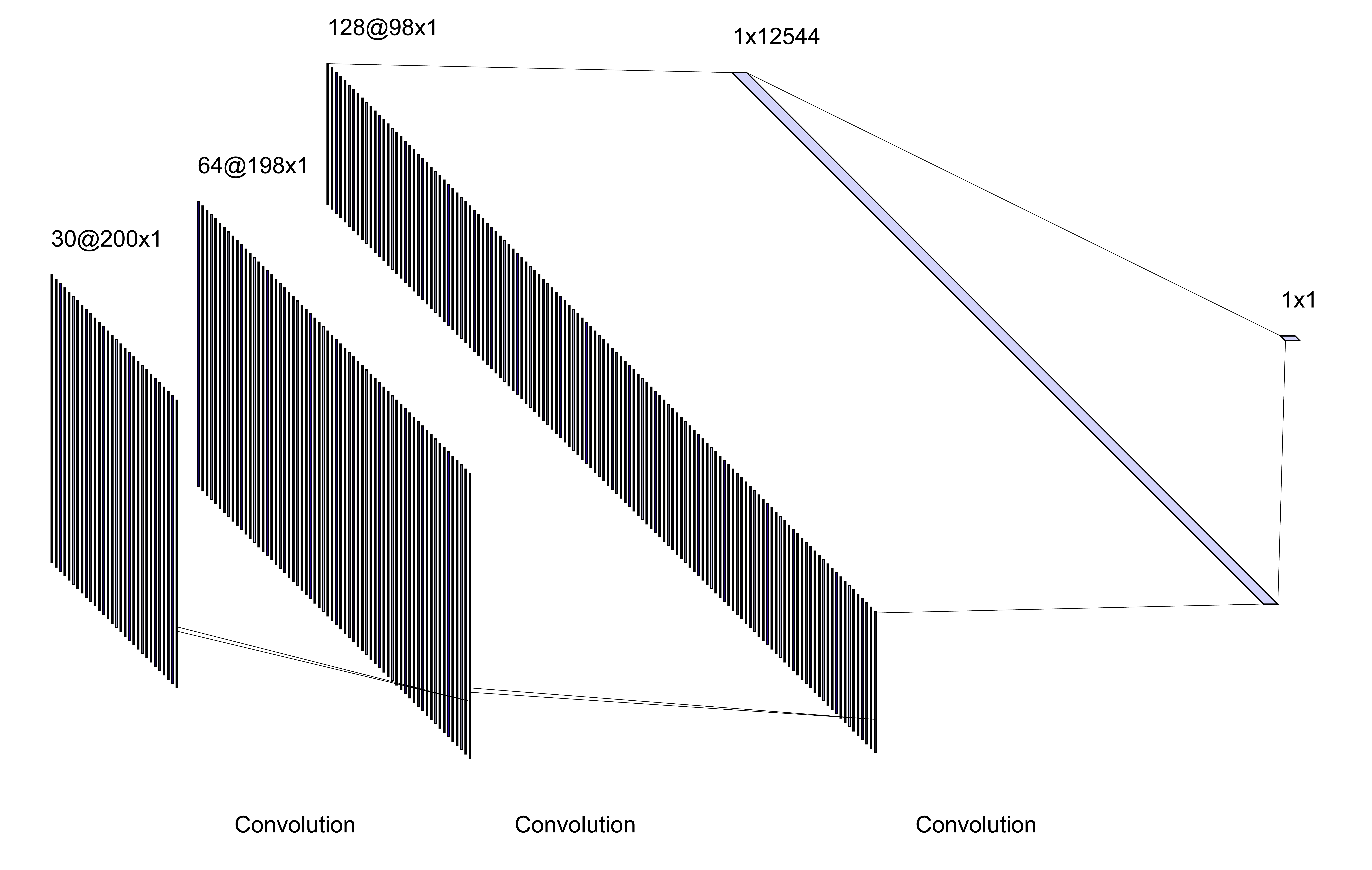}
        \caption{\textbf{1D CNN architecture for \textit{gradientless} data.}}
        \label{fig:arch-gradientless}
    \end{suppfigure}
    
    \begin{suppfigure}[H]
        \centering
        \includegraphics[width=\linewidth]{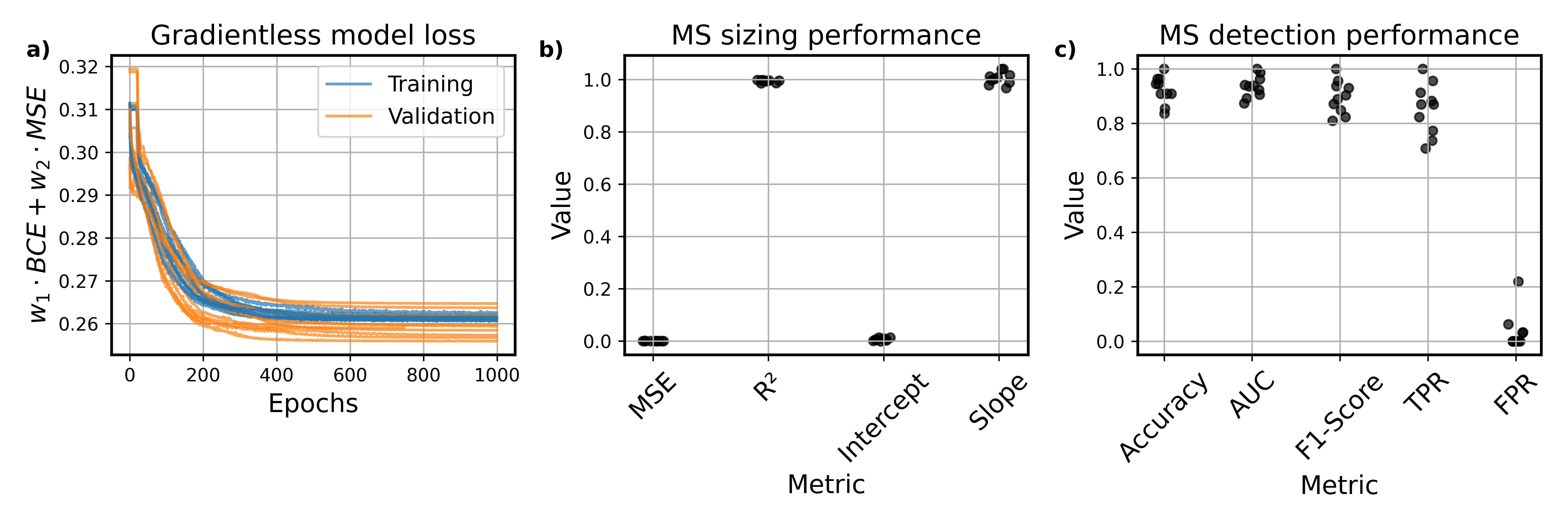}
        \caption{\textbf{Cross-validation results for \textit{gradientless} data.}\textbf{a) }Model loss during training and validation for 10-fold cross-validation. The consistent convergence indicates robust optimisation and generalisation across folds. \textbf{b) }Performance metrics for lesion volume estimation, showing the values across folds of the mean squared error (MSE), the goodness-of-fit coefficient $R^2$, intercept and slope values of the regression model between the simulated and the predicted MS volume. \textbf{c) }Performance metrics for the lesion detection, including accuracy, AUC, F1-score, TPR, and FPR.} 
        \label{fig:cv-gradientless}
    \end{suppfigure}
    
    \begin{suppfigure}[H]
        \centering\includegraphics[width=0.49\linewidth]{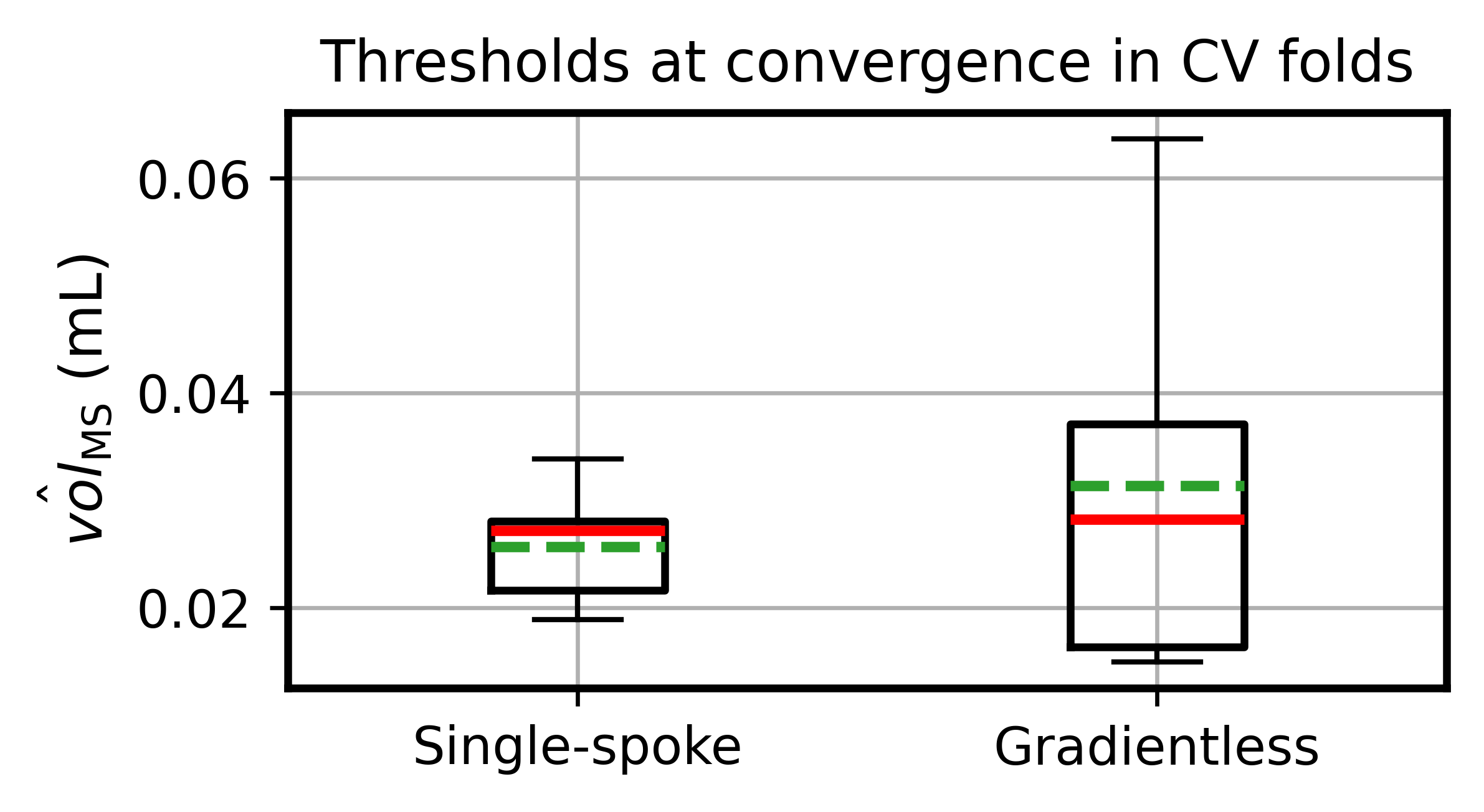}
        \caption{\textbf{Threshold selection for lesion classification at convergence across cross-validation folds.} The final threshold applied to the test set was the average of the ten cross-validation final thresholds (green dashed lines). The solid red lines indicate median values.}
       \label{fig:cv-thresholds}
    \end{suppfigure}

\subsection{Additional test set results}
\label{app:test_set_results}
    \begin{suppfigure}[H]
        \centering
        \includegraphics[width=\linewidth]{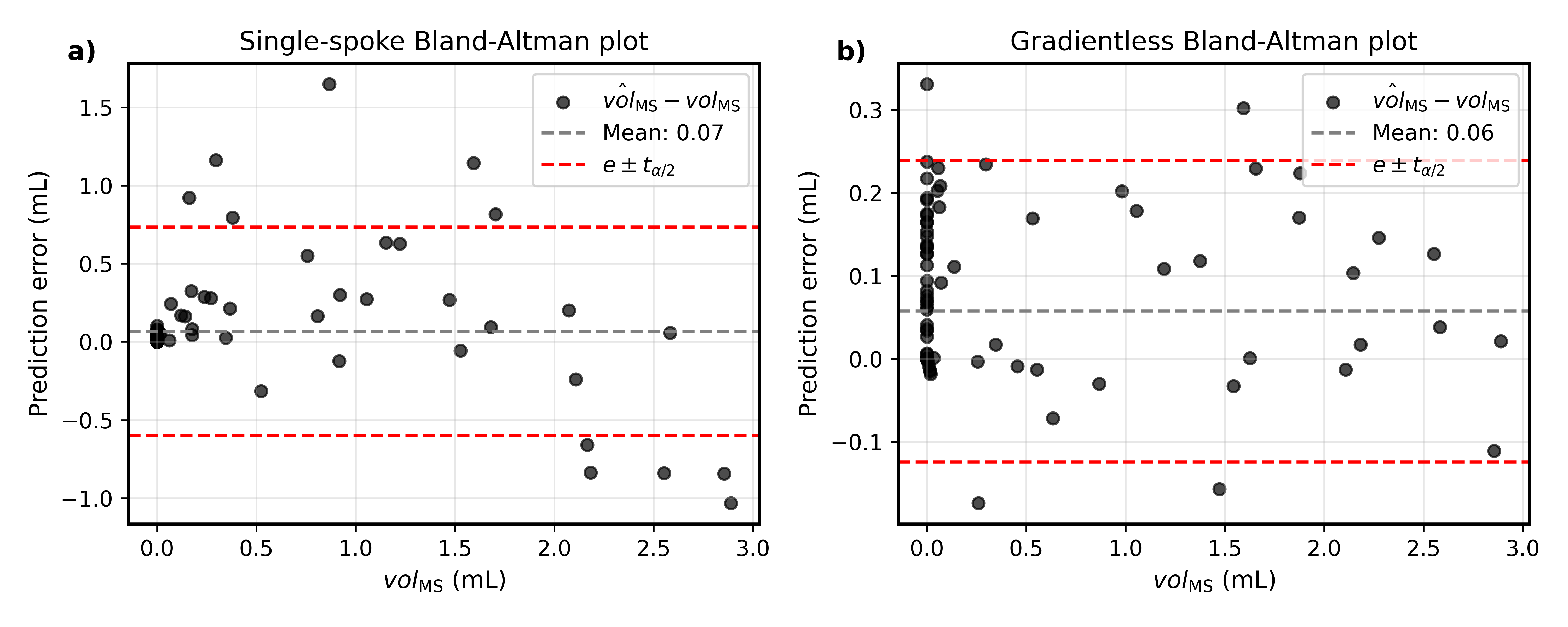}
        \caption{\textbf{Bland-Altman analysis for lesion volume estimation on the test set with the prediction error (bias) versus the simulated $\mathbf{vol_\text{MS}}$.}\textbf{ a)} \textit{Single-spoke} acquisition shows an increased mean bias of 0.07\,mL, with the limits of disagreement widened close to $\pm$0.5\,mL, indicating greater variability in prediction accuracy when spatial information is incorporated. \textbf{ b)} \textit{Gradientless} acquisition shows a mean bias ($\hat{vol}_\text{MS}-vol_\text{MS}$) of 0.06\,mL, showing most residuals within limits of agreement within $\pm$0.2\,mL for all $vol_\text{MS}$ values.}
        \label{fig:residuals}
    \end{suppfigure}

    \begin{suppfigure}[H]
        \centering
        \includegraphics[width=\linewidth]{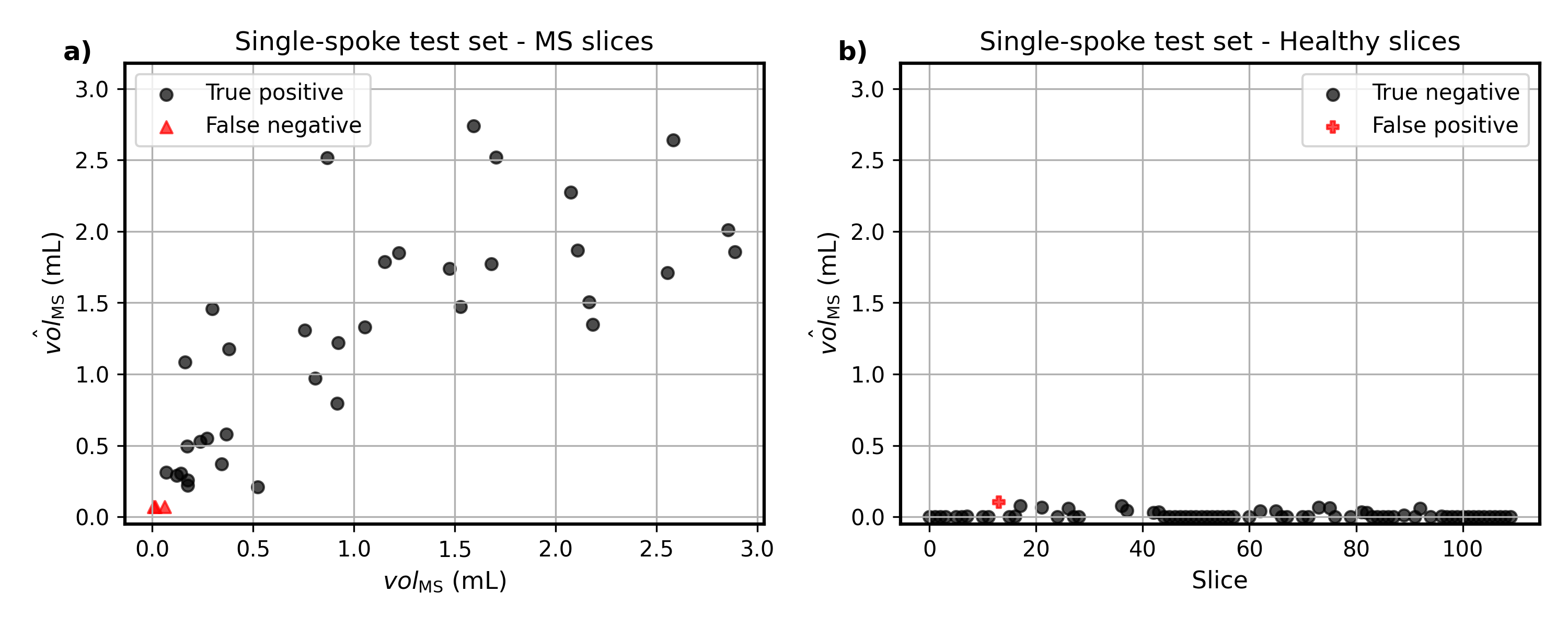}
        \caption{\textbf{Results with predicted ($\mathbf{\hat{vol}_\text{MS}}$) MS lesion volumes and final classification for \textit{single-spoke} data in the test set.} \textbf{a)} The scatter plot shows true positive predictions as black circles, while false negatives correspond to red triangles. Most false negatives correspond to low simulated volumes ($vol_\text{MS}\lessapprox0.0615$\,mL). \textbf{b) }The scatter plot shows true negative predictions as black circles, while the only false positive is represented with a red cross.} 
        \label{fig:class-single-spoke}
    \end{suppfigure}

    \begin{suppfigure}[H]
        \centering
        \includegraphics[width=\linewidth]{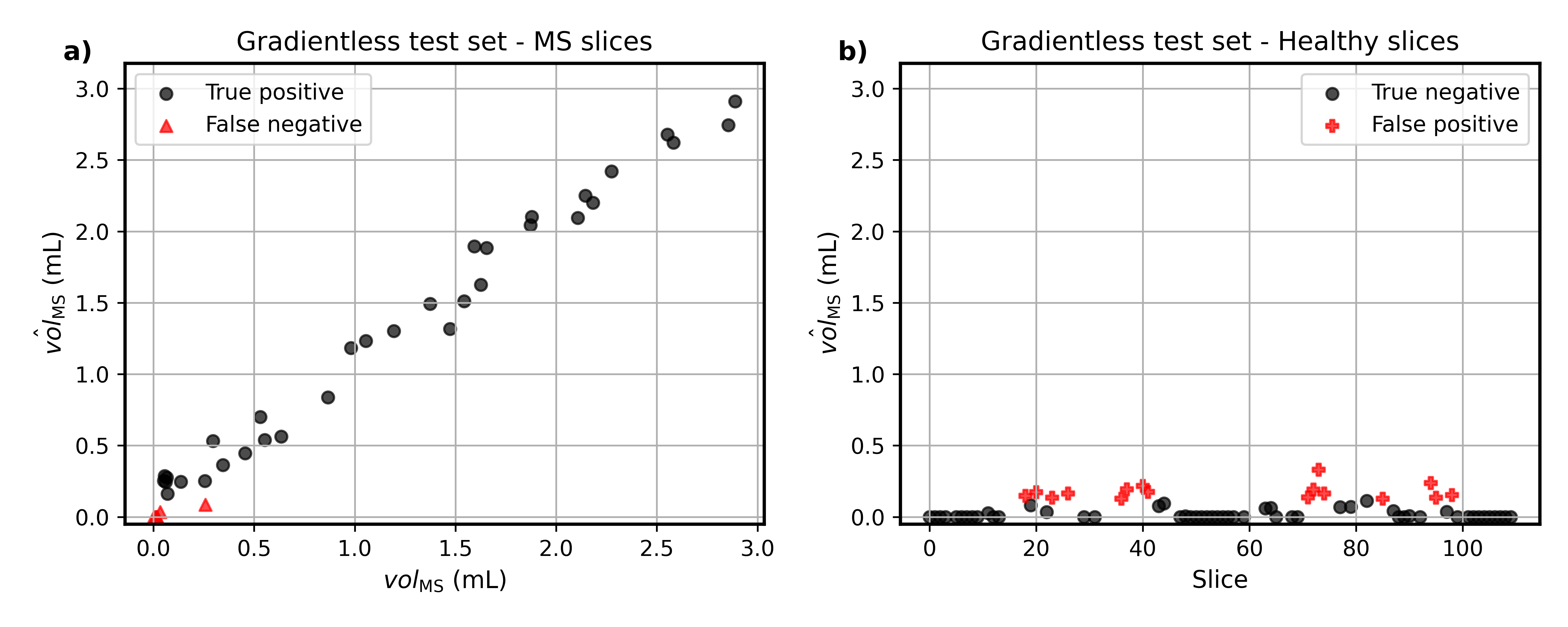}
        \caption{\textbf{Results with predicted ($\hat{vol}_\text{MS}$) MS lesion volumes and final classification for \textit{gradientless} data in the test set.} \textbf{a) }The predicted vs. simulated plot shows true positive predictions as black circles, while false negatives correspond to red triangles. Most false negatives correspond to low simulated volumes ($vol_\text{MS}\lessapprox0.04$\,mL), except for one outlying false negative, which corresponds to a lesion of $vol_\text{MS}\approx0.26$\,mL, reported in Figure~\ref{fig:testset-performance}. \textbf{b) }The scatter plot shows true negative predictions as black circles, while false positives correspond to red crosses. Most false negatives correspond to healthy slices with high average signal intensity, as reported in Figure~\ref{fig:zerospoke-healthyslices}.} 
        \label{fig:class-gradientless}
    \end{suppfigure}

     \begin{suppfigure}[H]
        \centering
        \includegraphics[width=0.6\linewidth]{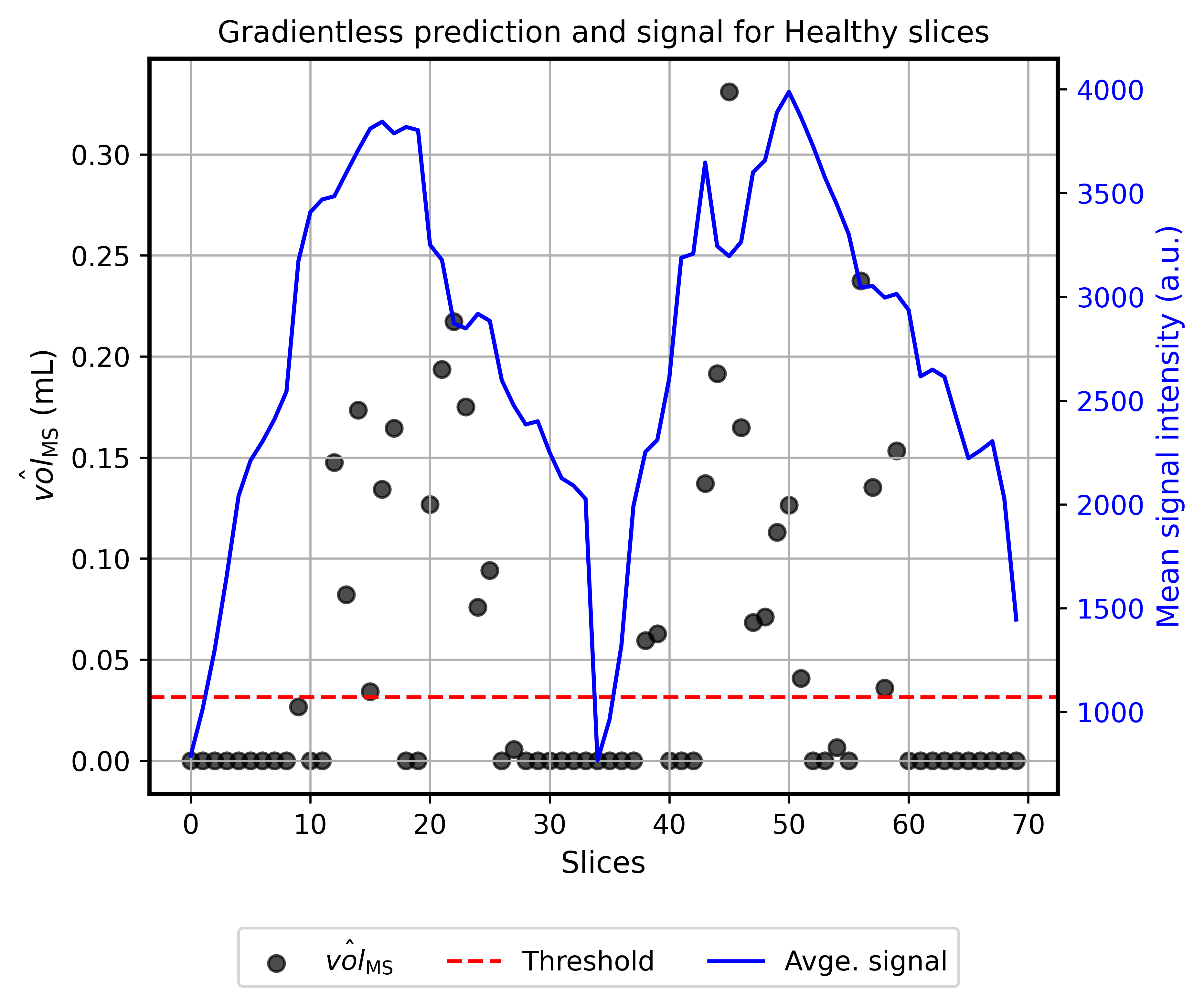}
        \caption{\textbf{Predicted ($\hat{vol}_\text{MS}$) for healthy slices in the \textit{gradientless} data test set.} The predicted vs. simulated plot shows true positive predictions as black circles, while false positives correspond to red crosses. The red dashed line corresponds to the prediction threshold for classifying slices as healthy or presenting MS. The blue lines correspond to the average signal value of each slice across the 30 TRs, represented in the right vertical axis. The plots show a correlation between the predicted MS volume and the signal intensity, which is related to the total tissue within the slice.}  
        \label{fig:zerospoke-healthyslices}
    \end{suppfigure}

    The performance of CNNs for the \textit{gradientless} acquisition, which yielded better outcomes regarding MS lesion volume prediction than the \textit{single-spoke} acquisition, was compared with two other techniques incorporating different amounts of a priori knowledge: ART and DE. Figure~\ref{fig:zerospoke-art-bbo} shows the results obtained with ART and DE for the \textit{gradientless} acquisition. Table~\ref{tab:performance_art_de} summarises the performance metrics of all techniques predicting the MS lesion volume. Besides, the same 1D CNN architecture was trained for a classification task, changing the loss function to the binary cross entropy and the final activation function to a sigmoid one. This CNN was included to investigate how a black-box model without a priori knowledge would perform if the supervised task used a binary response instead of a continuous one.  

     \begin{suppfigure}[H]
        \centering
        \includegraphics[width=\linewidth]{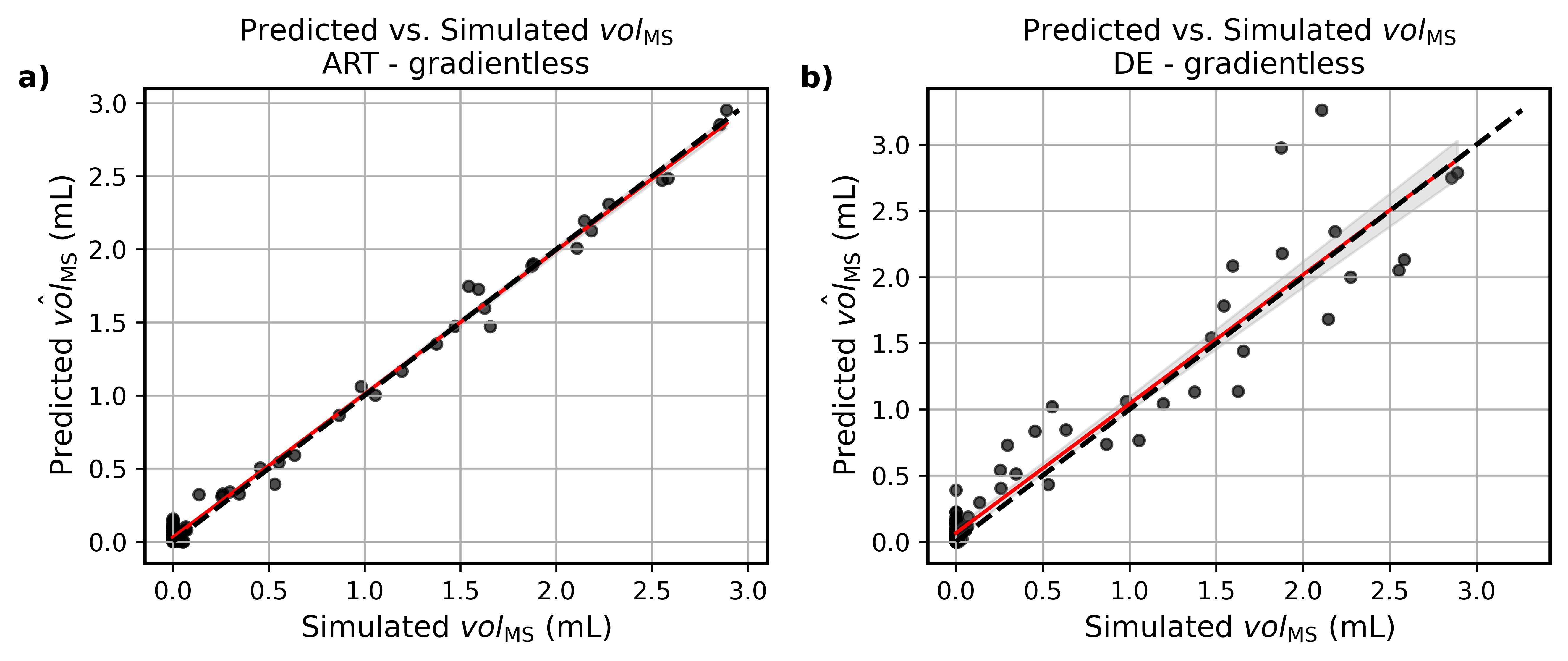}  
        \caption{\textbf{Validation of lesion volume estimation for \textit{gradientless} data with ART and DE.}\textbf{ a)} Predicted vs. simulated MS lesion volumes using ART. \textbf{b)} Same results using DE.}
        \label{supfig:zerospoke-art-bbo}
    \end{suppfigure}

    \begin{table}[!h]
		\centering
		\begin{tabular}{cccccccc}
        \toprule
			\textbf{Model} & $\mathbf{R^2}\uparrow$ & $\mathbf{b_0}\downarrow$ & $\mathbf{b_1}\uparrow$ & \textbf{AUC}$\uparrow$ & \textbf{TPR}$\uparrow$& \textbf{FPR}$\downarrow$ & $\mathbf{vol_\text{MS}^\text{FN}}\downarrow$\\\midrule
			\textit{ART} & 0.9943 & 0.043 & 0.9813 & 0.4875 & 0.975 & 1 & 0.05\\
                \textit{DE} & 0.9187 & 0.0956 & 1.0081 & 0.67321 & 0.975 & 0.6286 & 0.01\\
                \textit{1D CNN} & 0.985 & 0.0797 & 1.0081 & 0.7982 & 0.825 & 0.1571 & 0.26\\
                \textit{1D CNN (det.)} & -- & -- & -- & 0.8607 & 0.75 & 0.0286 & 0.26\\
			\bottomrule
		\end{tabular}
		\caption{\textbf{Model performance metrics in the test set for \textit{gradientless} acquisition without relaxation times variability.} Results are obtained against the same two phantoms in the test set. Upside ($\uparrow$) and downside ($\downarrow$) arrows indicate whether if higher (closer to 1) or lower (closer to 0) values, respectively, are better for each metric. See details in Table~\ref{tab:performance_metrics_test}. For both ART and DE, a single False Negative remained undetected, and, since the threshold on the predicted lesion volume was not optimised, it was set on $\hat{vol}_\text{MS} > 0$ to decide whether the slice presented MS lesions or not.}
	\label{tab:performance_art_de}
    \end{table}
    
    \newpage
    \subsection{Robustness tests}
    \label{app:robustness}
    Altered versions of the dataset were obtained to run three robustness tests. Differences in the signal-to-noise ratio (SNR) are a source of variability commonly encountered in MRI acquisitions due to hardware limitations, low magnetic field strength, or shortened acquisition times. This is particularly important for low-field portable MRI systems, which often operate with inherently lower SNRs. The effect of lower SNR was evaluated by progressively increasing the factor by multiplying the white Gaussian noise added to the $k$-space data in the preprocessing. SNR values ranged from 20 to \textit{``noise-less''}.
    
    Besides, apodisation was simulated to emulate shorter acquisitions in the \textit{gradientless} case and the loss of high-frequency information for the \textit{single-spoke} case. Information on low magnitudes at each TR's end was eliminated for the \textit{gradientless} acquisition. For the \textit{single-spoke} acquisition, information of high frequencies encoding finer details on the signal was gradually removed. Several percentages of low-magnitude or high-frequencies for each case were removed to obtain dataset versions with information loss. 

    As a last robustness experiment, we rejected the assumption that all slices (i.e., each ``patient'') share the same $T_1$ and $T_2$ relaxation times for each tissue. Instead, we drew $T_1$ and $T_2$ from distributions whose parameters were chosen according to reported values in literature \cite{larsson1988vivo}, approaching the physiological reality that relaxation times may vary across individuals. This experiment was run for the \textit{gradientless} simulation. Given its remarkable performance for MS lesion volume estimation, we considered it the best candidate to measure how the simulation of more realistic and variable data would impact model performance. Figure~\ref{fig:scatterplot_t1_t2_tissuevar} shows the bi-variate plot with the sampled relaxation times for all slices and tissues.  
    
    \begin{suppfigure}[H]
        \centering
        \includegraphics[width = \linewidth]{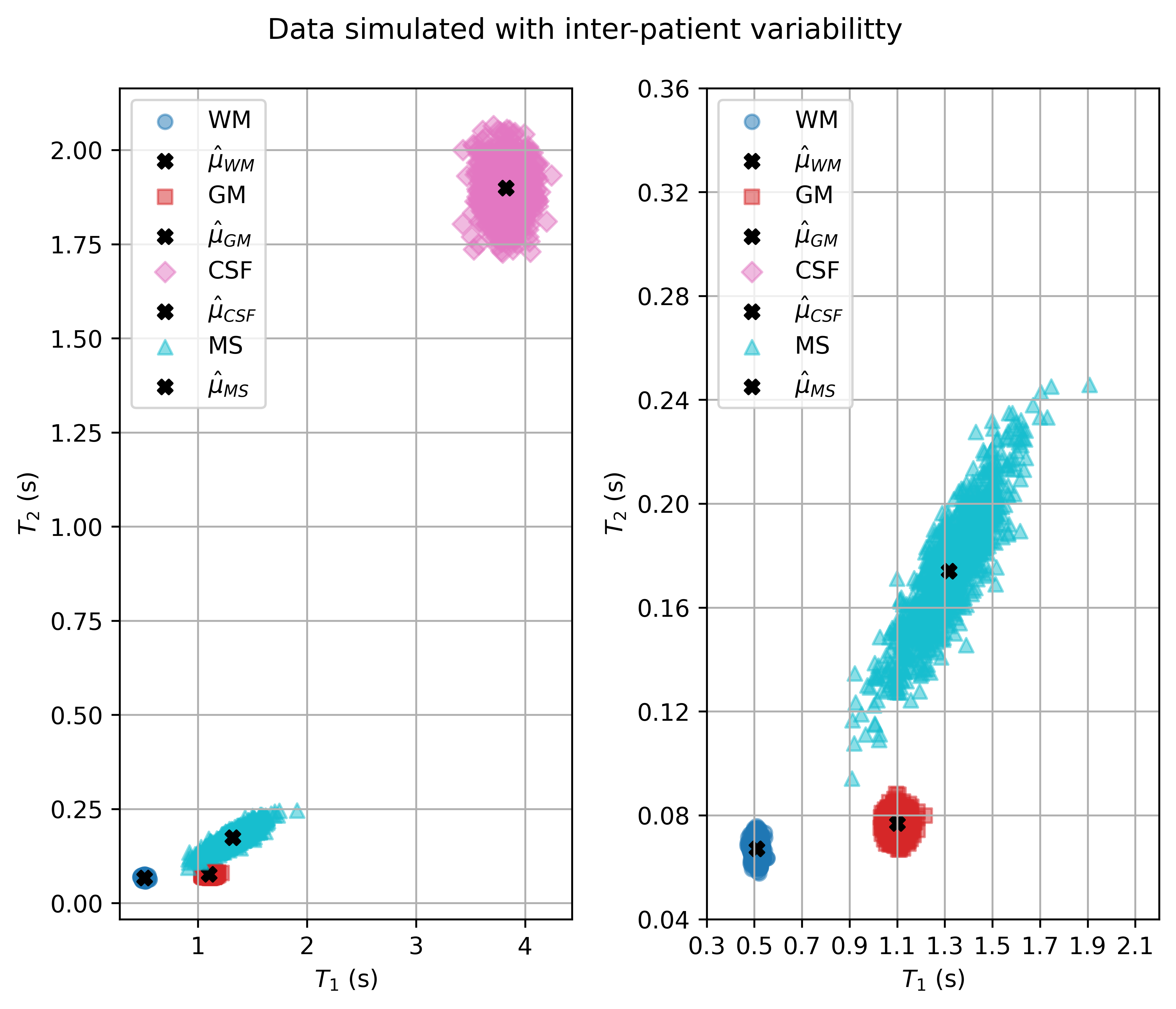}
         \caption{\textbf{Scatterplot of relaxation times $T_1$ and $T_2$ values for each slice with inter-patient variability.} Every marker corresponds to one slice, with each symbol and colour encoding a tissue type: white matter (WM), gray matter (GM), cerebrospinal fluid (CSF), or multiple sclerosis lesion (MS). The black crosses ($\mu_\text{WM}$, $\mu_\text{GM}$, $\mu_\text{CSF}$ and $\mu_\text{MS}$) show the mean values per tissue. The right panel is a zoom into the WM, GM and MS region. The average and standard deviation $(T_1; T_2)$ values (mean$\pm$std) are $(0.51\pm0.01 ; 0.067\pm0.003)$ for WM, $(1.10\pm0.03 ; 0.077\pm0.003)$ for GM, $(3.82\pm0.13 ; 1.90\pm0.06)$ for CSF and $(1.32\pm0.14 ; 0.174\pm0.023)$ for MS.}
         \label{fig:scatterplot_t1_t2_tissuevar} 
    \end{suppfigure}

    After training the same CNN architecture without variability on this dataset, we evaluated all three methods (ART, DE and CNNs) on the same two test phantoms, now with more heterogeneous slices. Table~\ref{tab:performance_interpat_var} summarizes the resulting performance. Because ART assumes known $T_1$ and $T_2$ values, mismatches caused by inter-slice variability appear to introduce large errors in ART's MS volume predictions,  with a decreased $R^2$ of 0.5311 (left plot in Figure~\ref{fig:zerospoke-art-bbo}). On the contrary, DE, which incorporates the estimation of relaxation times for each slice, showed a more robust and balanced performance than ART (right plot in Figure~\ref{fig:zerospoke-art-bbo}), with a solid $R^2$ of 0.8741. Finally, CNN-based approaches were also affected by the broader $T_1$ and $T_2$ ranges. The regression model was more impacted ($R^2$ of 0.7092 and AUC of 0.6939) than the detection one, which still kept an AUC of 0.7324 and a lower FPR compared to ART and DE, at the cost of missing some slices with lesions smaller than 0.37\,mL (lower right plot in Figure~\ref{fig:zerospoke-art-bbo}).

    \begin{suppfigure}[H]
        \centering
        \includegraphics[width=\linewidth]{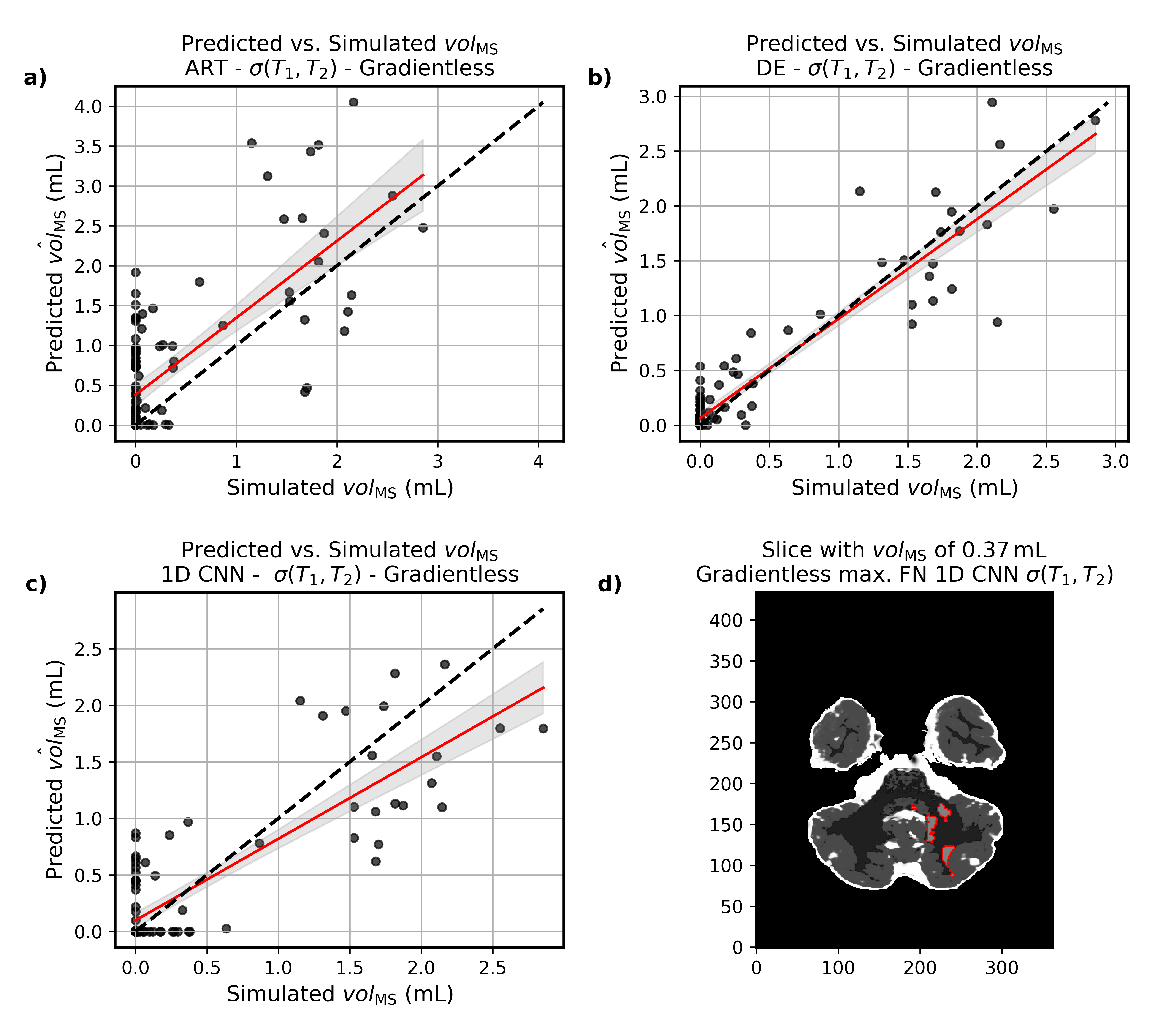}
        \caption{\textbf{Validation of lesion volume estimation with inter-patient $T_1$ and $T_2$ variability for \textit{gradientless} data.} \textbf{ a)} Predicted vs. simulated MS lesion volumes using ART. \textbf{b)} Same results using DE. \textbf{c) } Same results using the 1D CNN for regression. \textbf{d) } Slice with an MS lesion volume of 0.37\,mL which was the maximum undetected volume by the CNN. In all plots, the dashed diagonal represents the ideal $\hat{vol}_\text{MS} = vol_\text{MS}$ relationship, and the solid red line is a linear fit between simulated and predicted MS volumes.}
        \label{fig:zerospoke-art-bbo}
    \end{suppfigure}

    These results showed how performance can be downgraded by more realistic data, particularly affecting methods heavily relying on a priori knowledge (ART). This suggests that further work considering physics-informed models should include steps searching for parameter values governing such equations. Provided this search's good performance, models could benefit from this a priori knowledge, as suggested by DE results. On the other side of the spectrum, CNN-based models, being purely data-driven, were also impacted but were still able to find relevant data patterns within the same amount of drastically messier data. This also suggests that CNNs could benefit by increasing the training sample size via Data Augmentation by including additional physics-guided features in the CNN architecture, mitigating the impact of more variable data. 
    
\end{document}